# Music Viewed by its Entropy Content: A Novel Window for Comparative Analysis

G. Febres[1,2], K. Jaffé[2]

[1] *Departamento de Procesos y Sistemas, Universidad Simón Bolívar, Venezuela*
[2] *Laboratorio de Evolución, Universidad Simón Bolívar, Venezuela*

## Abstract

*Polyphonic music files were analyzed using the set of symbols that produced the Minimal Entropy Description which we call the Fundamental Scale. This allowed us to create a novel space to represent music pieces by developing: (a) a method to adjust a description from its original scale of observation to a general scale, (b) the concept of higher order entropy as the entropy associated to the deviations of a frequency ranked symbol profile from a perfect Zipf profile. We called this diversity index the '2nd Order Entropy'. Applying these methods to a variety of musical pieces showed how the space of 'symbolic specific diversity-entropy' and that of '2nd order entropy' captures characteristics that are unique to each music type, style, composer and genre. Some clustering of these properties around each musical category is shown. This method allows to visualize a historic trajectory of academic music across this space, from medieval to contemporary academic music. We show that description of musical structures using entropy and symbolic diversity allows to characterize traditional and popular expressions of music. These classification techniques promise to be useful in other disciplines for pattern recognition and machine learning, for example.*

# 1 Introduction

We all share the intuitive idea of music as a flow of ordered sound waves. Formally the presence of order in music was studied by Leonard Meyer [1], who pioneered the analysis of music as a phenomenon capable of creating emotions. Meyer analyzed in depth the expectancy experienced by the listener. In his explanations, Meyer used musical concepts and technical notations which are difficult to represent in quantitative mathematical terms. But the idea of music as a means to create specific sensations such as tension, sadness, euphoria, happiness, rest and completeness, is always present along his study. Meyer described the emotions caused by music as the result of the interaction between the sound patterns perceived and the brain. In his words [1]:

> *"The mind, for example, expects structural gaps to be filled; but what constitutes such a gap depends upon what constitutes completeness within a particular musical style system. Musical language, like verbal language, is heuristic in the sense "that its forms predetermine for us certain modes of observation and interpretation."† Thus the expectations which result from the nature of human mental processes are always conditioned by the possibilities and probabilities inherent in the materials and their organization as presented in a particular musical style." (†* Edward Sapir, "Language," Encyclopedia of the Social Sciences, IX (New York: Macmillan Co., 1934), 157.)

Meyer's referral to conditional probabilities implies the possibility of capturing some of the essence of musical style by observing the values of entropy associated with each music style. But the style of music has proved to be a difficult concept to handle. Similarly as occurs with other types of languages, the style is a way of classifying specific musical pieces. The determination of the style is based on characteristics describing the music, the time when it was composed, and the geographical context. Some researchers have set a style framework for music by quantifying those characteristics. In 1997 R. Dannenberg, B. Thom and D. Watson [2] produced readable Musical Instrument Digital Interface (*MIDI)* files by recording trumpet 10-second-long performances. Dannenberg et al used neural networks to classify the style of each recorded performance according to several features of music. In 2004 P. J. Ponce de León and J. M. Iñesta [3], measured the pitch, note duration, silence duration, pitch interval, non-diatonic notes, syncopation, and other music components, to build statistical characterizations of jazz and classical melody pieces. Perez-Sancho, J. M. Inesta and J. Calera-Ruiz [4] approached the same problem by categorizing the texts of music MIDI files. They extracted the melodies from the MIDI files and segmented the resulting texts into sequences of characters representing different lengths of music beats. In 2004, P. van Kranenburg and E. Backer [5] studied music styles starting from some music properties. They included the entropy of some parameters as properties. These studies indicate that it is possible to recognize properties related to the musical style in an automated fashion, but none fulfills the required generality as to be considered a true style recognizer. Music style is just a concept too fuzzy to serve as a quantitative reference framework useful to classify, with a single value, something as complex as music.

From a more theoretical perspective some researchers have provided useful schemas about the structures underlying music. In 2006 Mavromatis [6] presented models of Greek Chants depicting the melodic component of music as a process dominated by Markov chains. Later, in 2011 Rohrmeier [7] argues that that Markovian processes are too limited to properly model the complexity that arises when harmonies are added to melody. Rohrmeier proposes a Generative Theory of Tonal Harmony (*GTTH*) [7] as a set of recursive rules based on the Chomskian grammar and on the Generative Theory of Tonal Music (GTTM) by Lerdahl and Jackendorf [8]. Both branches of study, music as a phenomenon governed by Markovian processes, and the recursive context-free rules to model harmonies, are developed for music as it is written on the music-sheet. That is, music as an abstract entity represented by a set of meaningful symbols written on the music-sheet, which are supposed to represent the sonic effects pretended by the composer.

Even for this conception of music —its description on the music-sheet, which is simpler than recorded actual sounds— Rohrmeier points out in one of his notes in 2012 [9], that *GTTH* does not suffice to properly model the polyphonic music. On the other hand, in 2009 Mavromatis [10] suggested the application of the Minimal Description Length Principle (*MDL*) as an alternative to the Markovian models of melodies, and explained why *MDL* should be a powerful tool to describe music. Yet he announces these advantages are subject to the huge computational complexity foreseen of the algorithms associated to this type of analysis.

Even for the most intricate pieces of music, the music-sheet is rather simple when compared with the actual music and with the recorded file that can be reproduced —the sounds we hear. The quantitative analysis of music is even more demanding if polyphonic music is the subject of study. Polyphony adds more dimensions to an already almost unmanageable problem. To deal with polyphonic music Cox [11] measured the entropy of the sound for each time beat. Cox represents his results in two time-dependent entropy profiles: one for pitch and another for rhythm. The polyphonic music can be described as the superposition of many monophonic sound streams. The result is an overwhelmingly large number of combinations of sound frequencies. Luckily, all these sound streams are synchronized in time and therefore its record in a file leads to a one-dimensional text, where some character sequences may form patterns that represent the musical elements contained in the text-file.

Working independently, Febres and Jaffe [12] presented the Fundamental Scale Algorithm (*FSA*). A method based on the *MDL* Principle applicable not only to music, but to most problems in which the recognition of patterns in a large string of written symbols, is an issue. The *FSA* is capable of unveiling the 'dominant' symbols of a description. In the present work we apply the *FSA* to 453 MIDI files containing academic, traditional, and popular music. For each piece, the Fundamental Symbols —the set of symbols leading to the description minimal symbolic entropy— was determined, and the symbol frequency profiles built. In order to compare the shape of profiles based on different number of symbols, a method is devised and presented. Additionally, a measure of Higher Order Entropy and a method for its calculation, is proposed. We used these methods to represent different types of MIDI-music in an entropy-diversity space. The dependence between the type of music and the selected representation-space is analyzed.

## 2.     Methods

Understanding the structures underlying music is an old restlessness, always present among researchers During the 50's, Meyer [1] and, more recently, Huron [13] have linked the music structure to our emotions and expectations. Their description of musical structure and its influence in our emotions is based on considerations of explicit musical language, as written on the music sheet. By using other analytical resources, a group of researchers, Mavromatis[13] among them, offer models for the construction of melodies that assume that a Markovian process is behind each specific style of melody. These models, based on Finite State Machines (FSM) generalized in stochastic terms by a Hidden Markov Model (HMM) [14], are able to produce melodies that fit into a certain music style, after being properly trained. Extending the HMM to include harmonies, requires the identification of an inconveniently large number of states. As an alternative method Rohrmeier [7] proposes a system of grammar rules to model harmonic progressions, an important extension of Lerdahl and Jackendorf [8] previous work and their Generative Theory of Tonal Music (GTTM).

Music can be seen as a recursively nested group of structures (Rohrmeier [15]). Even considering just melody, music consists of kinds of fractal structures leading any attempt for its analysis, to a very complex task. Attempting to model polyphonic music 'amplifies' these difficulties so much, that Rohrmeier [15] considers the analysis of the structures of polyphonic music, a practically impossible task.

In those studies where the focus is on the music sheet, the analysis is limited to the music as the composer intended it to sound —instruments, rhythms and tempo, scales, note pitches, keys, chords, temperament, volume, etc.—, but leaving out of the assessment of many other effects of real music which are present when it is performed with musical instruments. This study, on the contrary, is done with the recording of sounds as expressed in computerized music files. Subtleties as the effects of relative position of the instruments, their timber, syncopation, mistuning, the performer's style and even errors, are represented in these files, up to some degree depending on the recording quality and resolution.

## 2.1. Language Recognition, Diversity and Entropy

In this study we propose a radically different method to study the structure of music. Instead of analyzing the symbols written on the music sheet, we look at the sound recorded from an actual performance, by reading the text associated to the computerized file which contains the recording. To do this, we inspect the sequence of characters of the computerized files viewed as texts.

Even for short files, this is not a simple task. A music file read as a text, is a long sequence of characters which does not exhibit recognizable patterns, resulting in a code extremely difficult to interpret. Not knowing the rules of a grammar system it is not possible to decide a priori how to recognize the symbols needed to interpret the description. There are no words in the sense we are used to, and the characters we see do not indicate any meaning for us. We cannot even be sure about the meaning of the space character " ". Thus music files contain character strings to represent sounds according to the coding system used and the selected discretization level. But, as opposed to natural language text files, the music files do not show words or symbols that we humans can recognize without the help of some decoding device. Therefore, to find some order within these symbols —sequences of characters— that are camouflaged with the surrounding text, we consider the entropy of each possible set of symbols, that is, each possible way of reading the same written message.

We claim that the set of symbols whose frequency distribution corresponds to the lowest (or is near to), possible symbolic entropy value, is a good representation of the structure of the language used for the description. We call this set the Fundamental Symbols, and the method used to its determination is the Fundamental Scale Algorithm [12]. The result is the set of symbols $Y_i$ which can reproduce the description with such a frequency distribution $P(Y_i)$, that the entropy associated with, is minimal. The set grouping the Fundamental Symbols is regarded as the Fundamental Language $B_*$. The asterisk as sub-index is used to recall that $B$ is the result of an entropy minimization process. Thus we can write

$$B_* = \{Y_1, \ldots, Y_i, \ldots, Y_D, P(Y_i)\}. \tag{1}$$

In Expression (1) the diversity —the number of different symbols— is represented as $D$. Once we know the set of Fundamental Symbols along with the frequency of each fundamental symbol —equivalent to their probability distribution—, we can compute its symbolic specific diversity and the entropy of each piece of music, applying Expressions (2) and (3). The specific diversity $d$ is calculated as

$$d = \frac{D}{N}, \tag{2}$$

Where $D$ is the diversity of language $B$ —the number of different symbols in the description— and $N$ is the total number of symbols, repeated or not. A version of Shannon's entropy $h$ [16], generalized for languages comprised of $D$ symbols, is used to compute quantity of information describing each music piece. The probabilities of occurrence of symbols $Y_i$ are the components of the one-dimensional array $P$:

$$h = -P \, \log_D P. \tag{3}$$

## 2.2. Frequency Profiles

The value of entropy *h* is a good base for the comparison of descriptions. But it may be not be enough to properly represent the many dimensional differences of entities as those we are dealing with. For that reason, we complement our treatment of each music piece with the shape associated to the values of array **P**. To obtain the shape, we ranked the symbols according to their appearance frequency **P(Y$_i$)** and plotted **P(Y$_i$)** vs *Rank(Y$_i$)*, both in logarithmic scales; pretty much as the method to build a Zipf's profile.

### 2.3. Higher Order Entropy

The use the symbolic diversity *d* and the symbolic entropy *h* to characterize music pieces, allowed for differentiating among genres of music. However, two pieces of music, even though having different ranked frequency profiles, may share similar values of entropy. When this is the case, the difference between two profiles can be described as the way they 'oscillate' around their respective middle line. Thus, we looked at the patterns of these oscillations, and quantified them by computing the value of the entropy associated with these oscillations. Therefore, elevating the comparison of descriptions to a finer level of detail. Details of the mathematical formulation to compute the Higher Order Entropy are shown in Appendix B. In this work we use the 2$^{nd}$ order entropy, and we refer to it with the symbol *h* and adding a superscript between brackets (*h$^{[2]}$*).

### 2.4. Scale Downgrading

As mentioned, we used the shapes of frequency profiles as characteristic of a music piece. When comparing the shape of several frequency profiles, the different number of symbols for which each profile was created, is a problem. To solve this we present a method we called Scale Downgrading, useful to represent a symbol frequency profile with a smaller number of symbols while keeping it general shape. Details of the mathematical formulation to compute the Scale Downgrading are shown in Appendix C.

## 2.5. Music Selection

Music is the result of the superposition of a vast variety of sounds. But music sounds respond not only to the information written on the music sheet, but also the addition of small differences introduced by the interpreter. Music is then the result of a large number of different symbols to form sounds sequences. These sound sequences are included in the file produced by the recording of a musical piece. In spite of the unreadable condition of any of these files for us humans, the files contain all the information regarding the music, and thus we can appreciate this information as music when we reproduce the file and hear it. Due to limitations of the Fundamental Scale [12] Algorithm and the enormous complexity of most conventional music recording formats, we had to rely on file MIDI coding to discretize the symbols forming these pieces of music while keeping the computations within a feasible condition for our algorithm in its current condition. Using formal music recording formats as .MP3, .MP4 or .WAV, is still desirable and a matter of further improvement of technical aspects of the Fundamental Scale Algorithm [12]. Yet, MIDI-music provides the conditions for us to advance with this study.

Most MIDI files include metadata at their beginnings and their ends, usually written in English or Spanish. The length of these headers and footers can be considered small compared to the total symbolic description length; since cleaning all files would represent a non-automated task, we decided not to prune this small amount of noise and leave the files as they show when opened with a .txt extension.

Table 1 shows a synthesis of the music selection we used as subject to apply the entropy measurement method. The selection includes pieces from classical and popular music of different genres. Our music library is organized in a tree. To have some reference of the place where a music piece, or group of pieces, is located within the tree, we assigned a name to each tree level. Table 1 shows this classification structure fed with more than 450 pieces from 71 composers and 15 different periods or types of music.

## 3. Results

All pieces of music were organized in a classification tree to which we refer to as *MusicNet*. By computing the Fundamental Scale to all leaves of *MusicNet*, we were able to obtain the fundamental symbols of each music piece included in our dataset, as well as for each music subset defined by composer, type, genre, period, or any other characteristic property of the included music. *MusicNet* is too lush to be extensively presented here. But we include the upper levels of the tree in Table 1 and a link that allows access to the whole tree in Appendix A. Table 1 displays the datasets of MIDI music used for our tests and values of specific diversity, entropy and 2$^{nd}$ order entropy accompanied with their respective standard deviations.

**Table 1.** Music classification tree *MusicNet*, and the data associated to top levels of the tree.

| | | | | | | | Spec. diversity | | Entropy | | 2nd Ord. Ent. | |
|---|---|---|---|---|---|---|---|---|---|---|---|---|
| Class | Type | Period/Style | Region | Genre | Composers | Pieces | Ave. | Std.Dev. | Ave. | Std.Dev. | Ave. | Std Dev |
| | | | | Total | 71 | 453 | | | | | | |
| Western | Academic | Medieval | | | 12 | 40 | 0.062 | 0.026 | 0.649 | 0.048 | 0.949 | 0.037 |
| | | Reinainssance | | | 10 | 31 | 0.048 | 0.016 | 0.622 | 0.037 | 0.935 | 0.041 |
| | | Baroque | | | 8 | 55 | 0.039 | 0.013 | 0.581 | 0.057 | 0.911 | 0.050 |
| | | Classical | | | 7 | 45 | 0.040 | 0.019 | 0.566 | 0.059 | 0.896 | 0.049 |
| | | Romantic | | | 13 | 89 | 0.049 | 0.021 | 0.602 | 0.068 | 0.914 | 0.061 |
| | | Impressionistic | | | 4 | 34 | 0.050 | 0.015 | 0.582 | 0.052 | 0.921 | 0.044 |
| | | 20th Century | | | 8 | 35 | 0.052 | 0.017 | 0.559 | 0.057 | 0.888 | 0.062 |
| | Traditional | | Venezuelan | Traditional | >20 | 56 | 0.049 | 0.014 | 0.540 | 0.056 | 0.929 | 0.036 |
| | Popular / Contemp. | | | Movie Themes | | 18 | 0.048 | 0.010 | 0.615 | 0.051 | 0.934 | 0.033 |
| | | | | Rock | 5 | 24 | 0.041 | 0.010 | 0.585 | 0.043 | 0.919 | 0.045 |
| | | | | Jazz | | | | | | | | |
| | | | | Regie | | | | | | | | |
| | | | | Tecno | | | | | | | | |
| Asian | Traditional | | Hindu-Rag | Raga | Several | 14 | 0.083 | 0.019 | 0.697 | 0.061 | 0.974 | 0.026 |
| | | | Chinese | | Several | 12 | 0.048 | 0.015 | 0.582 | 0.038 | 0.915 | 0.046 |

## 3.1 Diversity and Entropy

Diversity and entropy are quantitative characterizations of communication systems. Within the scope of a communication system, the diversity and the entropy may reveal differences regarding style or even period of its evolution. All pieces of our music library are organized in three groups: occidental academic, traditional and Rock/Movie Themes. Diversity vs. length and entropy vs. length graphs are shown in Figures 1 and 2.

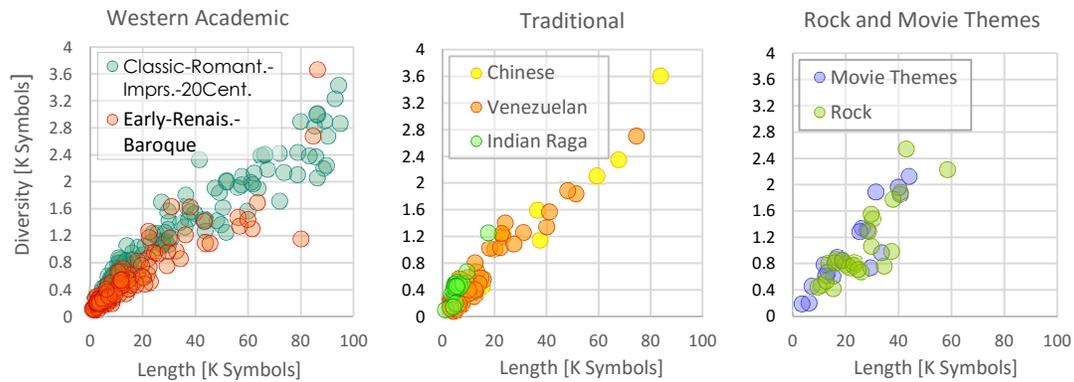

**Figure 1.** Diversity as a function of piece length measured in symbols for different classes of music.

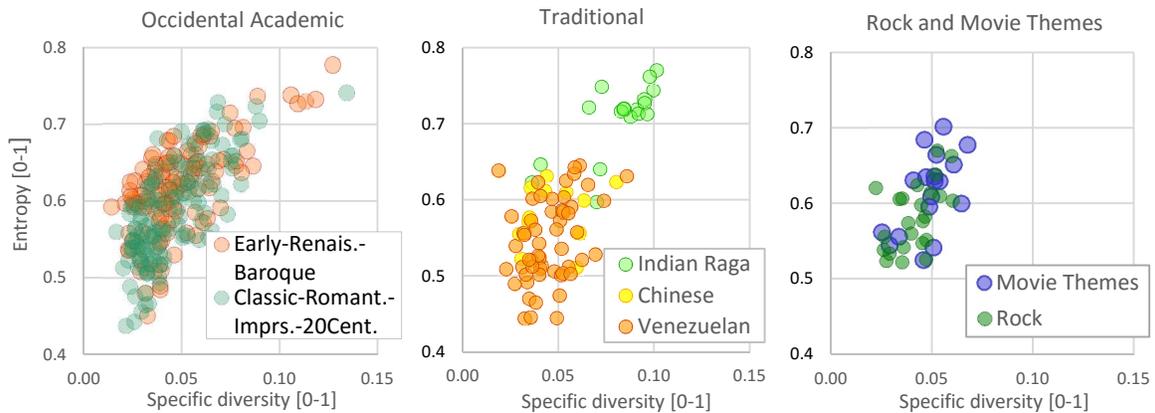

**Figure 2.** Entropy as a function of specific diversity for different classes of music.

## 3.2 Information Profiles

We are interested to know the effect of degrading the scale of observation of musical descriptions. Prior to this computation, we know that degrading the scale —equivalent to viewing the system from a remoter perspective—, means observing less details, therefore, as the number of symbols used in the description decreases, we expect to get less information. Thus, there are at least two reasons to inspect these information profiles: (a) to evaluate if they capture information about the music's type or class. (b) To obtain a sense of the minimal degraded diversity that maintains the essence of the system, by showing a shape that resembles the description at its original symbol diversity. Using this *minimal degraded diversity* allowed us to compare the shapes of many music frequency profiles at the same diversity; a condition needed for a fair comparison.

We built the information profiles for several pieces of music. To obtain them we started from the description at their original symbol diversity $D$, and degraded the observation scale $S$ by applying Equations (C3), (C4a), (C4b) and (C4c). An example of information profile is presented in the bottom of Figure 4. This information profile was built with the amount of information (the entropy) corresponding to the symbol probability profiles observed at different downgraded scales (shown in the upper section of Figure 4). The downgraded values of the diversity were selected, so that at any scale, the number of degrees of freedom of the symbol frequency profile (symbolic probability profile) is a power of 2. The number of degrees of freedom of any probability distribution is $k-1$, being $k$ the number of different categories in the distribution. Thus, the number of different symbols considered for each degraded symbol diversity is $S = 2^i + 1$, where $i$ is a positive integer.

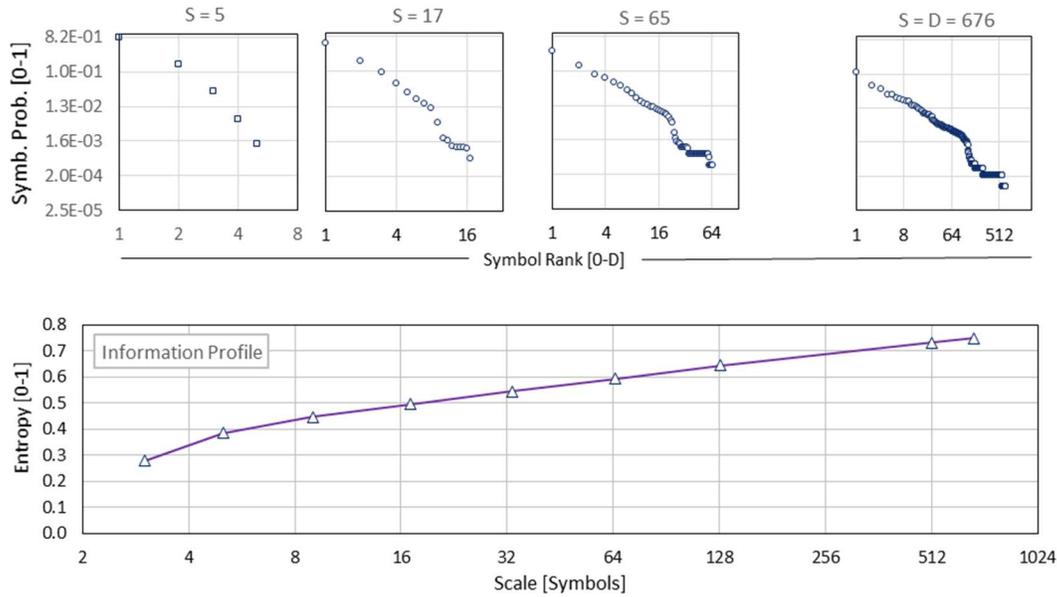

**Figure 3**. Variation of frequency profiles for several degraded scales and Information profiles calculated for *Hindu-Raga.Miyan ki Malhar*.

Two additional information profiles are presented in Appendix D. When comparing the information profiles at different scales for the *Hindu-Raga.Miyan ki Malhar* with the other two music pieces, it is visually clear that, the Hindu-Raga piece differentiates showing a promontory in the profile at a diversity *S=17*, that none of the other present at that scale. But the downgraded diversity *S=17* is not detailed enough to recognize the slight differences between the profiles of *Beethoven.Symph9.Mov_3* and *LAURO.Antonio-ValsVenezolanoNro3.Natalia,* included in Appendix D. In order to find visually different profile shapes among the three samples analyzed, we had to inspect the profiles with a diversity *S=129*. With that level of refinement in the profile drawing, we were able to distinguish each music pieces' profile from another; we thus selected this diversity value (*S=129*) as the diversity we should downgrade all pieces in order to obtain characteristic property values for each piece.

## 3.3 Symbol Frequency Profiles

A way to visualize the differences between two classes of music is to draw the ranked symbol frequency profile. Each profile has *D-1* degrees of freedom. That means the profile's shape can be altered in *D-1* different ways by modifying the frequency of the *D* different symbols which make the musical piece description.

Figure 4 shows the symbol frequency profiles computed for our sample of impressionistic music; Graphs (a) and (b) show first and second order symbol profiles correspondingly. Frequency profiles computed for all the groups of music contained in our data set are included in Appendixes F and G. All frequency profiles in Appendix F were computed at a scale or downgraded diversity *D=129*, using the numeric values of the probability of each symbol and each style of music, which are included in Appendix F. The 2$^{nd}$ order frequency profiles shown in Appendix H were all computed at a downgraded diversity *D=33*.

When observing these frequency profiles, a reasonable question arises: Are these profiles capable of depicting the organized change that might be produced by an evolution process of music? The seven graphs corresponding to academic music, from Medieval to 20$^{th}$ Century music, suggest that the answer is *yes*. For most periods and music styles, the frequency profiles exhibit two easily recognizable regions: a higher ranked frequency region located toward the head of the ranked distribution, and a second region at the right of the ranked distribution, which extends until the symbol rank's cut-off value where sometimes an elbow shaped profile appears near the last ranked symbol at rank = *D =129*. For Medieval music, the distribution head's region occupies most of the profile range, showing a bow shaped profile. While the academic type of music covers the time until the 20$^{th}$ Century period, this bowed section progressively shortens until the transition of the two regions reaches the middle of the logarithmic horizontal axis. The last tail elbow also softens till it disappears at the classical music profile. The slope at the transition zone also shows a gradual increase from the medieval music, where transition zone is very soft, up to the 20[th] Century music, which shows a rather stiff transition zone. The vertical range of the profiles also grows as the time period progresses; with the only exception of Impressionistic music, all other considered styles of academic music, require a larger range of different frequency values in the vertical axis when compared with its previous music period.

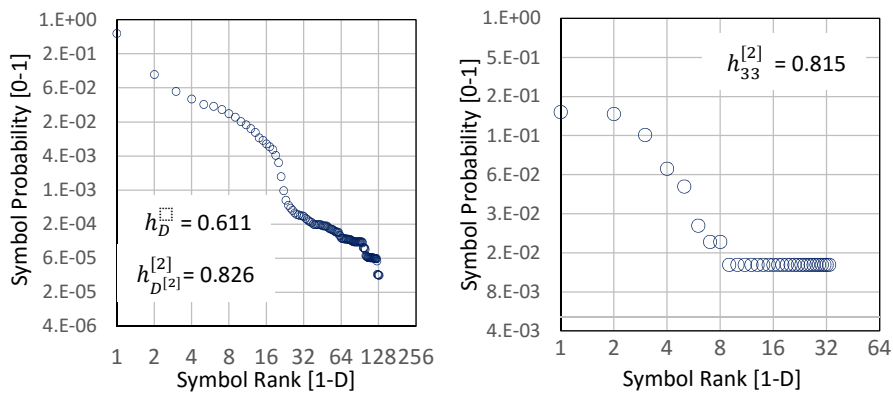

**Figure 4**. Symbol ranked frequency profiles for registered impressionistic music. Graph (a. Left) shows the traditional symbol profile. Graph (b. Right) shows the profile for the 2[nd] order symbols.

When looking at traditional and popular music, we observe a shorter vertical range of values if compared against the academic music profiles. From all non-academic music considered, Hindu-Raga music exhibits the flattest profile while Chinese music has the steepest one.

The comparison of these profiles suggests that it is possible to capture structural music differences by observing these shapes. On the other hand, there are profile similitudes between some pairs of classes of music. Baroque music and Rock music, for example, have similar shaped profiles. Also, music from Impressionistic and Chinese periods, exhibit similar overall profiles. However, reducing the profile shapes down to a quantifiable index proves to be difficult and perhaps over simplistic. In this sense, the inclusion of an additional characteristic, as is our recently defined 2[nd] Order Entropy, seems to be justified.

### 3.4  Clusters and Tendencies

The frequency profiles built lead us to obtain values of symbolic diversity, entropy and 2nd order entropy for our selected set of MIDI-musical pieces. These values are shown in Appendix E. To extend the observation space up to a three-dimensional space, we present 3D graphs representing the values of Appendix E. Figure 5 presents 3D graphs for the diversity $d$, entropy $h^{[1]}$ and 2nd order entropy $h^{[2]}$ of our music data sample.

In the graphs of Figure 5 each bubble corresponds to a single music piece. When a musical work is complex and can be divided by parts —for example: suites, concerts and symphonies—, each part is considered a single piece and is represented by a bubble. Figures 6 and Appendixes J and K show the average values of the same properties, but this time, computed for sets of musical pieces grouped according to Music type and composer, thus, in these Figures each bubble corresponds to a different music period/styles or a composer. Three views of the same 3D plot are presented.

Figure 6 reveals how all periods of academic music locate in different sectors of the 3D space formed by diversity $d$, entropy $h$ and 2nd order entropy $h^{[2]}$. It is worthwhile to mention that in Figure 6 the size of the bubbles do not represent the dispersion of the music pieces grouped under a music style or period, thus there is more overlapping among types of music than that suggested by the representation of the bubbles in the graph.

Returning to Figure 5, the Graphs shown may appear, at first glance, as a disorganized mix of bubbles representing music styles in our 3D space. Certainly, there are clusters of types of music sharing the local space. Therefore it would be difficult to split some clusters according to their location. However, in spite of the difficulty to see through this dense cloud of bubbles, for some specific types of music, the separation of their cluster's locations seems a feasible task. Medieval music (old-rose bubbles), for example, occupies a subspace of relatively high entropy and high diversity if compared to the location of Renaissance music (light green bubbles). Following the chronological time direction, Baroque music (dark green bubbles) maintains the general tendency towards a reduction of its symbolic diversity $d$ and its entropy $h$ (represented as $h1$ in the 3D graphs). Comparing Classical music (light blue) with Baroque, its predecessor in time, we observe a stabilization of diversity $d$ and entropy $h$ values, however, there appears a noticeable reduction in the values of the second order entropy $h^{[2]}$ (represented as $h2$ in the 3D graphs). On the other hand, if we consider 'distant' types of music as Hindu-Raga (yellow bubbles), and Venezuelan music (orange bubbles), there is very little, or none overlapping between the spaces where the bubbles are; these clusters occupy different spaces and our representation allows separate them.

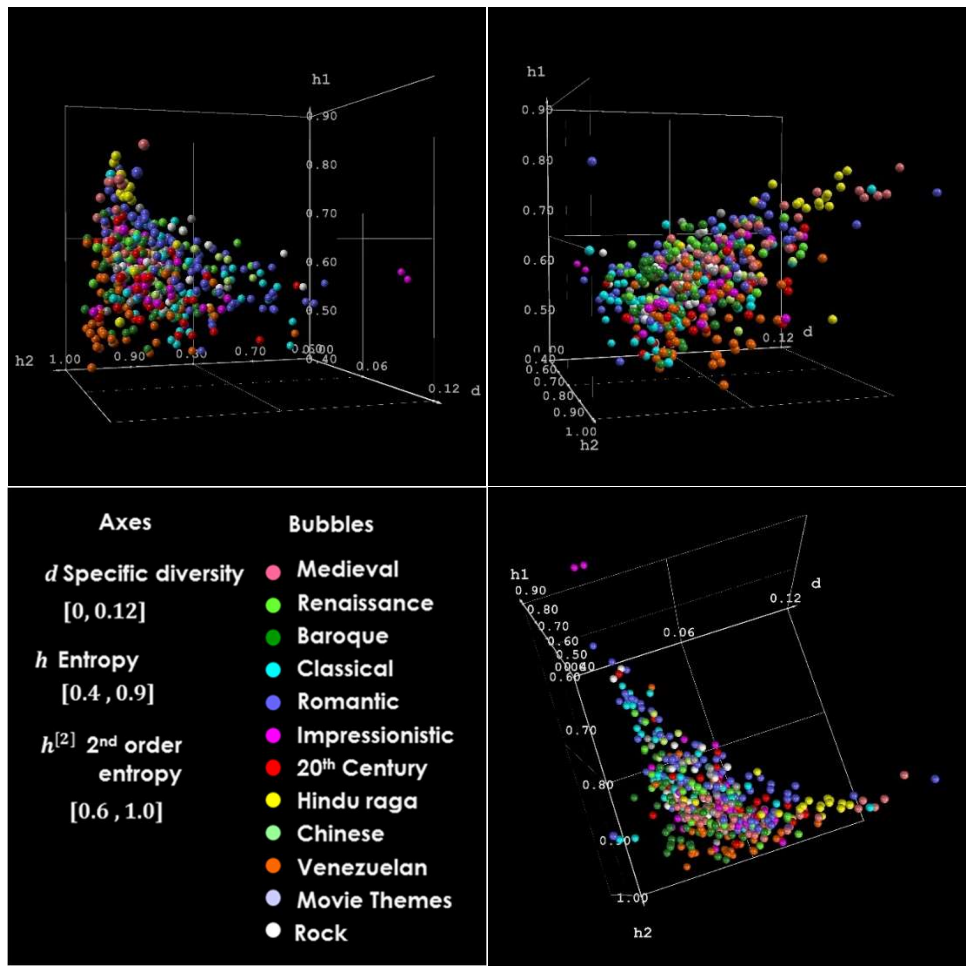

**Figure 5:** Three views of the representation of MIDI-music selected pieces in the space specific diversity, entropy, 2nd order entropy ($d$, $h^{[1]}$, $h^{[2]}$). Each bubble represents a MIDI-music piece.

To have a quantitative sense of the overlapping occurring for these clusters, we present the averages and standard deviations of the properties that characterize each type of music in our sample. Tables 2 and 3 show the results. We address some aspects of this discussion in Section 4.

To appreciate any tendency of specific diversity *d* and entropies *h* and *h[2]* over time, we plotted these variables as functions of time. The resulting graphs are included in Figure 7 and Appendix I. For Chinese and Hindu-Raga music pieces we do not have information about the time when they were composed. We, therefore, did not include those types of music in these graphs.

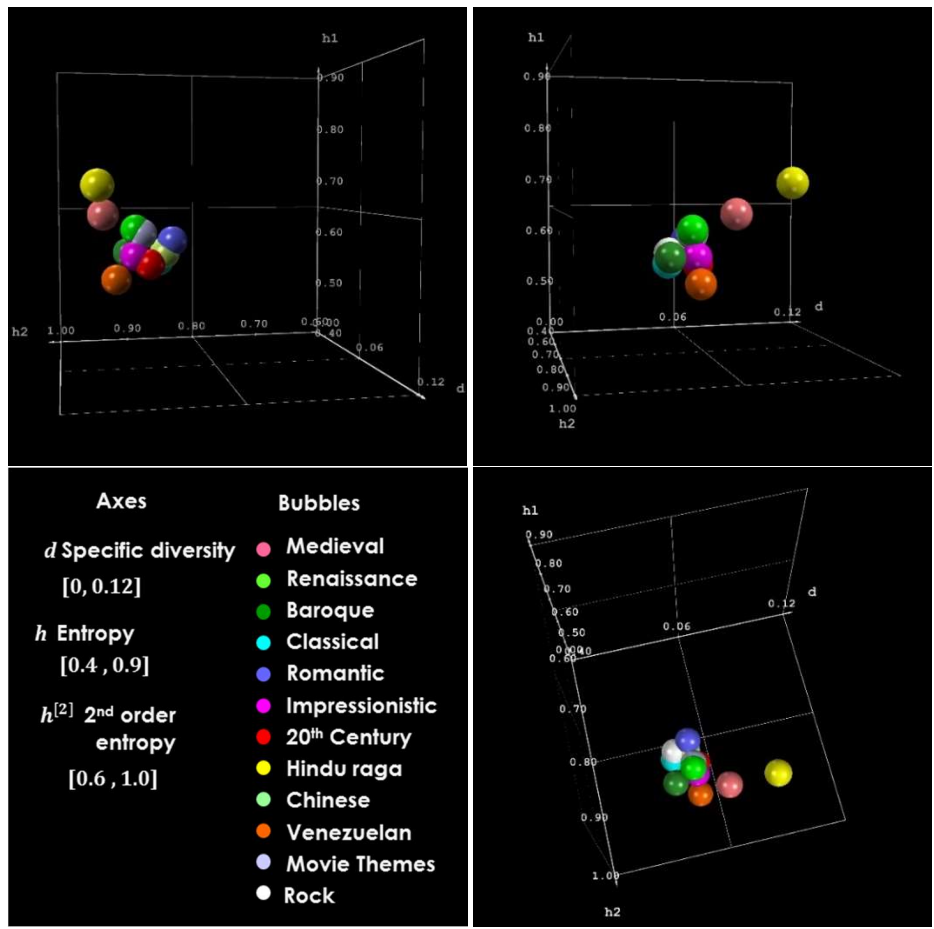

**Figure 6:** Three views of the representation of music period/style groups in the space specific diversity, entropy, 2nd order entropy (*d, h[1], h[2]*). Each bubble represents a group of music pieces sharing the same style/period.

**Table 2:** Properties of western academic music.

| | | Medieval | Renaissance | Baroque | Classical | Romantic | Impress. | 20th Century |
|---|---|---|---|---|---|---|---|---|
| | Num.Elem. | 40 | 31 | 55 | 89 | 45 | 34 | 35 |
| Specific diversity $d$ | Average | 0.0618 | 0.0479 | 0.0388 | 0.0403 | 0.0485 | 0.0500 | 0.0518 |
| | Std.Dev. | 0.0258 | 0.0159 | 0.0127 | 0.0190 | 0.0210 | 0.0150 | 0.0168 |
| Entropy $h$ | Average | 0.6489 | 0.6219 | 0.5806 | 0.5661 | 0.6023 | 0.5819 | 0.5592 |
| | Std.Dev. | 0.0475 | 0.0373 | 0.0566 | 0.0592 | 0.0676 | 0.0521 | 0.0570 |
| 2nd order entropy $h_{[2]}$ | Average | 0.9446 | 0.9014 | 0.9085 | 0.8664 | 0.8521 | 0.8829 | 0.8917 |
| | Std.Dev. | 0.0320 | 0.0629 | 0.0499 | 0.0700 | 0.0945 | 0.1153 | 0.0679 |

**Table 3:** Properties of some traditional and popular music.

| | | Hindu Raga | Chinese | Venezuelan | Movie Thms. | Rock |
|---|---|---|---|---|---|---|
| | Num.Elem. | 14 | 12 | 56 | 18 | 24 |
| Specific diversity $d$ | Average | 0.0828 | 0.0476 | 0.0493 | 0.0485 | 0.0415 |
| | Std.Dev. | 0.0189 | 0.0153 | 0.0143 | 0.0104 | 0.0103 |
| Entropy $h$ | Average | 0.6971 | 0.5818 | 0.5398 | 0.6150 | 0.5853 |
| | Std.Dev. | 0.0607 | 0.0380 | 0.0558 | 0.0511 | 0.0431 |
| 2nd order entropy $h_{[2]}$ | Average | 0.9539 | 0.8608 | 0.9259 | 0.8915 | 0.8577 |
| | Std.Dev. | 0.0288 | 0.0777 | 0.0614 | 0.0104 | 0.0706 |

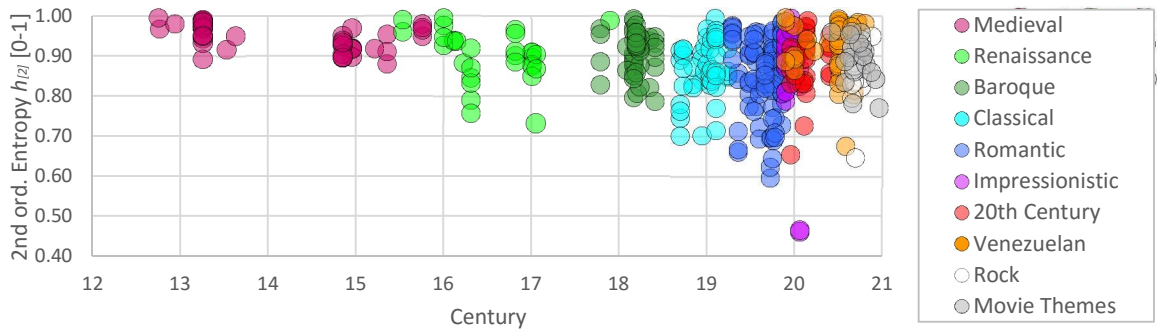

**Figure 7:** Variation of 2nd order entropy over time for several types of music.

# 4 Discussion

Music can be transmitted by sounds and by writing. But, the communication of music by writing lacks of its essence and does not produce, at least not for most people, the emotions and sensations associated to a pattern of sounds. Music writing shall be considered as a useful tool for composing, making arrangements, recording, and teaching music. Thus, transferring musical information is possible by means of music sheets or other kinds of music written representation. But, we think that rigorously speaking, written forms of music convey information, instead of conveying real music. Thus we devote our discussion to some of the properties of the information associated to a set of MIDI-music files. Our purpose is to demonstrate that the properties of these texts, even though in an indirect manner, can be used to characterize, and up to some degree to depict, the actual pattern of sounds that we call music.

## 4.1 Diversity and Entropy

The dependence of Diversity $D$ vs Length $N$ is nearly linear. Only for short music pieces, the Diversity-Length curve shows slight concavity. For all other ranges, the Diversity $D$ of music can be modelled as linear relationship with the length $N$ of the music description. The slope change observed near the origin, may be due to the English and Spanish overhead texts, which are generally included to start and to end the MIDI files. These natural language segments are considered as noise and its presence should not have an important affect over the overall music description when the music piece is reasonably large in terms of symbols. Nevertheless, the specific diversity $d$, represented by slope $D/N$, keeps close to a constant value for every type of music, becoming a characteristic value that may distinguish one type or style of music from another. Figure 1 illustrates how the point clusters for different types of music, tend to group around different lines, leading to different averages of specific diversity $d$ as shown in Tables 2 and 3. The value specific diversity measured for individual pieces ranges from 0.0183 (*Academic:*

*Impressionistic: RAVEL.Maurice, Bolero2*) to 0.1341 (*Academic: Romantic: SAINTSAENS.Camille: CarnavalDesAnimaux: 08.PersonnagesLonguesOreilles*). Complete set of values can be found in the link signaled in Appendix A.

As can be seen in Figure 3, the graphs show that entropy is aligned to a very stiff slope in the space entropy $h$ vs specific diversity $d$, and even though the entropy values represented, fill a wide range of values (from *0.45* to *0.8*), they seem to closely follow an average curve of the form $h = d^{\alpha}$, similar to those found for human natural languages in a previous work [17]. The large dispersion of entropy is then a consequence of the small range of specific diversities $d$ where music establishes. Nevertheless, the values of the entropy standard deviation observed in Tables 2 and 3 are, in general, small compared to the range of entropy averages, suggesting that entropy values capture some of the essence of the type or period of music and therefore justify its inclusion in a music entropy model. Values of the $2^{nd}$ order entropy average go from *0.89* (*Academic: $20^{th}$ century*) to *0.97 (Asian: Traditional: Indian Raga*). The standard deviation is about *0.05* and in general smaller than the range of variation of the average $2^{nd}$ order entropy from one group to another.

## 4.2 Frequency Profiles

Figure 4 include graphs with the information needed for a system description, in its vertical axis, and the scale of observation —the diversity $D$ of symbols used in the description— in its horizontal axis. These graphs have been called with two different names; researchers who consider Shannon's information [16] as a direct measure of complexity [18,19] call it *Complexity Profile*. Those who consider complexity as the pseudo-equilibrium [20–23] that the system reaches when it bounds its disorder by self-organizing its symbols, prefer to call these graphs *Information Profiles*. These names, which refer to the same type of graph, arise from the different interpretation of complexity. The first group of researchers see complexity as proportional to the length of the symbolic description, while the latter group pays more attention to the system's activity to keep itself organized. Despite the fact these names refer to different concepts, both seem to be valid, Any case, these information profiles show how sensitive are the description lengths of a MIDI-music piece to the change in the observation scale, represented here by the downgraded diversity.

We traversed two paths for our procedure. In a path we inspected the shapes of the ordered frequency profiles for all types of music included in this study. By visually comparing them, we found similarities between the profiles of different types of music; Baroque and Rock showed very similar shaped profiles, as well as the chronologically successive periods, Romantic and impressionistic, also did. We also found that Hindu-Raga and Venezuelan music have the flattest and the steepest profile shapes respectively, locating their shapes at opposite extremes of a scale somehow built to evaluate these shapes.

### 4.3 About the Evolution of Music

Indeed, in Figure 5 it turns out difficult to distinguish the dominant locations for all types of music. The locations of individual pieces of some types of music are dispersed and their central location is not easily recognized. But this does not mean that a piece, properly classified as a specific type of music, does not lie relatively near a certain location which corresponds to the type of music in the space considered. Take for example Romantic music (darker blue bubbles) and Baroque music (darker green bubbles). Despite the noticeable dispersion of the bubbles, each group occupies a different volumes within the space represented in Figure 5. Both music-type clusters are shaped as bows. The one representing Romantic music, is located toward the center of the cube, while the Baroque music-type cluster is located near the high second-order entropy corner. This aspect of the discussion is important because the standard deviations of the three properties evaluated, presented in Tables 2 and 3, are of the same order as the variation of the averages of the same properties, thus giving the false idea that these clusters are indistinctly dispersed all over the same space and are, therefore, not separable by the properties suggested here. The reason why our clusters may be separable while exhibiting high standard deviations, with an apparent full overlapping, resides on the shape on the clusters; they are not spherical, instead they are shaped as thick arched sheets.

Figure 6 shows how each type of music tend to occupy different sector of the space diversity-entropy. Focusing in the academic music, it can be seen a progressive move from the location of Medieval music, located in the sector of high diversity and entropy, to the location of more recent music as the Classical and Impressionistic, located at relatively lower specific diversity and entropy. The ordered locations of each type of academic music upon the time parameter, suggests that some types of music evolve in a way that can be detected in the mentioned space; $d, h^{[1]}, h^{[2]}$).

Hindu traditional raga and Venezuelan traditional music are easily recognizable. There must be some properties that make them well defined and different from each other. The fact Hindu-Raga and Venezuelan music appear far from any other style of music in Figure 6, does not surprise. On the contrary, it should be taken as sign of goodness of the space $d, h^{[1]}, h^{[2]}$ to represent music differences, and confirming the prominent distinctions between the profile shapes seen for these types of music in the profiles shown in Appendixes F and G.

Considering Graphs in Figures in Appendix G, we see that academic music has evolved to produce profiles associated with a lower value of the 2nd order entropy; for academic music this tendency seems sustained from the medieval music to the impressionistic period. Traditional and Popular music exhibit a 2nd order entropy comparable to the academic 20th Century's music.

The specific diversity *d*, on the other hand, reveals a slight reduction with time but an increase of dispersion of this variable, shown starting from the classical music and the romantic period, does not allow us to make a clear statement about the sustained tendency of a reduction of the specific diversity over time. On the side of traditional and popular music, specific diversity and entropy show less dispersion than their counterpart from academic music at comparable times.

Figure 7 and Appendix I show the behavior of variables *d*, $h^{[1]}$, and $h^{[2]}$ each one plotted vs time for academic music. There seems to be a tendency to lower the value of those variables with time. But the evident increase of the dispersion of these indexes, hides the overall change over time of academic music's entropy. Yet, when the three properties *d*, $h^{[1]}$, and $h^{[2]}$ form a joint view and time is a parameter, a clustering migrates from an extreme position to another emerges from the graphs (Figure 5 and 6). Suggesting that the combination of the properties *d*, $h^{[1]}$, and $h^{[2]}$ offer a good basis to build a space where the music style can be recognized.

## 5  Conclusions

The texts produced with music coded by the MIDI synthesizer, are susceptible to be analyzed using symbolic diversity and entropy as variables which can be used to characterize music type, and even more subtle properties, as style. The inclusion of higher order entropies accentuates the detectable differences between music styles.

We did not use any knowledge of the mechanisms of the MIDI coding process.  We started looking at file texts that seemed to be totally meaningless and not decipherable. Discovering the set of fundamental symbols for each music text description we found several important facts: 1: There is a fundamental symbol set that describes each piece of music. 2: The Fundamental Scale concept, presented in former works, is useful for determining the set of fundamental symbols of machine-coded texts, as MIDI-music text descriptions. 3: The scale downgrading method proposed allows for comparison of properties of systems of different nature and at different scale.

By applying the Fundamental Scale Algorithm, we have gone beyond the theoretical considerations about the Minimal Description Length Principle. We built frequency symbol profiles which work as quantitative descriptions for several hundreds of MIDI-music pieces. Due to the shapes of these profiles, which are practically unique, these profiles represent a sort of 'signature' of the complete polyphonic sound of each musical piece, with all its subtleties and complexity. After comparing our results for musical pieces according to their music style and period of time, we can affirm that the method works as a consistent procedure to visualize and to classify music styles and to quantify differences among them. Due to the shapes of the clusters representing each type of music, which are far from-spherical, we did not attempt to create probability fields for each type of music in the space diversity-entropy. However, we foresee the possibility for handling transformations to the shape of the space $d$, $h^{[1]}$ and $h^{[2]}$, to achieve the conditions required for a reasonable separation of these clusters, or alternatively, to estimate the probability associating each location in that space with each type of music. But that would lie within the scope of a future work. For the time being, locating text descriptions in the space specific diversity, entropy and 2nd order entropy, presents as a promising tool for classifying MIDI-music descriptions, with applications in many research fields as quantitative linguistics, pattern recognition, and machine learning.

Music is a reflex of social and cultural likes. We have strived to compare music styles over a quantitative basis. Our results reveal that for all the indexes used to characterize musical genres and styles, there is an increasing dispersion over time; perhaps the image of a society constantly committed to overcome any cultural barrier, thus making music an expanding phenomenon which grows in any direction of the space we use to observe it. This novel quantitative way of analyzing music might eventually allow us to gain a deeper insight into the musical structures that elicit emotions, illuminating the working of our brains and getting a better handle on music.

## 6. Acknowledgments

We want to thank those musicians and enthusiasts who produce several web sites where MIDI sequences of several kinds of music are available, accompanied with additional information presented in well-organized sites. Here we explicitly acknowledge them: midiworld.com, gregorypino74.jimdo.com/partituras-instrumentales, classicalarchives.com/midi.html, cse.iitk.ac.in/users/tvp/music/, faculty.pittstate.edu/~ilia/music/Mchinese.html, venezuela.ch .

## Appendix A. Music property tree. *MusicNet*

MIDI MUSIC PROPERTIES.
**http://www.gfebres.net/F0IndexFrame/F132Body/F132BodyPublications/MusicComplexityModels/MusicNet.htm**

## Appendix B. Higher order entropy

For an ordered symbol frequency distribution, entropy can be used as a general concavity –or convexity– profile index. To obtain an indication about the oscillations of the profile around the middle line represented by the Zipf's distribution reference line, a new index must generated. We propose the entropy of the distance between the distribution profile and the Zipf's reference as the new index. Figure 1 illustrates the basis for the definition of this new entropy level.

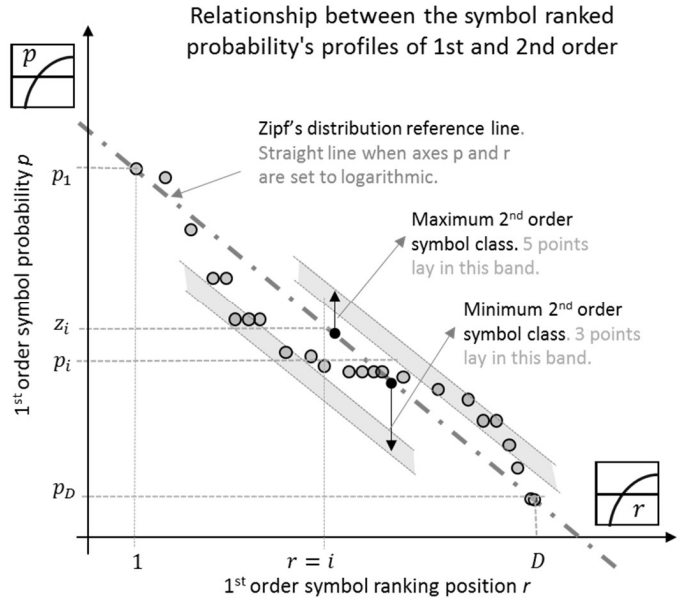

**Figure B1**. Typical symbol ranked probability profile with examples of 2nd order symbol bands. Each dot represent the probability of finding a symbol within all the symbols forming a system description. 1st order symbols are ranked according to their probability of appearance. The most common symbol appears in first place ($r = 1$) and the least frequent symbols appear at the end or tail of the ordered probability distribution representation ($r = D$).

To differentiate these two entropy calculations, we will call this *first order entropy,* or simply, entropy. We refer to the newly created concept as the *second order entropy*. For an ordered probability distribution profile, its first order entropy of is sensitive to its overall shape. Since any change of the profile slope needs to run along a wide range of the horizontal axis in order to impact the weighted area calculation, local changes in the profile slope are not effectively captured with the entropy. Second order entropy, on the contrary, is an index that focuses in the gap between the ordered symbol frequency distribution and the reference Zipf's distribution, it senses therefore the shape of the oscillations of the symbol probability profile.

To obtain a measure sensitive to small oscillations –or slope changes– we focus the distance $E$ between the symbols probability and the imaginary perfect Zipf's distribution $z_i$ that best fits the profile subject to study. The distribution $z_i$ is calculated as follows:

$$z_i = \frac{k}{i^g} \quad , \quad g = \frac{p_1 - p_D}{D} \quad , \qquad \text{(B1a) (B1b)}$$

Where $g$ is the Zipf's distribution slope and $k$ is a real number to stablish the starting point on the Zipf line for the first ranked symbol. Notice that $k$ is not necessarily equal to $p_1$, as is usually presented. Here the value of $k$ have to be adjusted to lead to a unitary area under the Zipf's distribution. The distance $E_i$ between a symbol probability $p_i$ and the imaginary Zipf's distribution $z_i$ is presented as a one-dimensional array.

$$\boldsymbol{E} = \begin{bmatrix} E_1 \\ E_2 \\ \vdots \\ E_D \end{bmatrix} = \begin{bmatrix} p_1 - z_1 \\ p_2 - z_2 \\ \vdots \\ p_D - z_D \end{bmatrix} . \tag{B2}$$

As depicted in Figure 1, the size of these deviations around the Zipf's profile can define a new language: the *second order language*. To obtain the 2nd order language we need to define the smallest $\boldsymbol{E}_{min}$ and the largest $\boldsymbol{E}_{max}$ and a resolution $q$ to establish the size of the bands to classify the symbols between the values of $\boldsymbol{E}_{min}$ and $\boldsymbol{E}_{max}$. After some arithmetic, these band boundaries can synthetized as the one-dimensional array $\boldsymbol{B}$ as:

$$\Delta q = \frac{E_{max} - E_{min}}{q} \quad , \quad B_i = B_{i-1} + \Delta q \quad , \quad B_1 = E_{min} - \frac{\Delta q}{2} \qquad \text{(B3a)(B3b)(B3c)}$$

$$\boldsymbol{B} = \begin{bmatrix} B_1 \\ B_2 \\ \vdots \\ B_q \end{bmatrix} \tag{B3d}$$

Vectors and distributions associated to an order $u$ are represented using a supra-index enclosed by squared brackets. The transition matrix $\boldsymbol{U}$ to relate the distribution at order $u$ with the distribution at order $u-1$, is represented using the supra-index formed by the supra-index *[u, u-1]*. The symbol probability distribution associated to the 2nd order language is represented by array $\boldsymbol{P}^{[2]}$ and obtained as indicated by Expression (12).

$$\boldsymbol{P}^{[2]} = \boldsymbol{U}^{[1,2]} \cdot \boldsymbol{P}^{[1]} . \tag{B4}$$

$$\boldsymbol{U}^{[1,2]} = \begin{bmatrix} U_{1,1} & U_{1,2} & \cdots & & U_{1,D} \\ U_{2,1} & \ddots & \cdots & \cdots & \vdots \\ \vdots & \vdots & U_{i,j} & & U_{i,D} \\ & \vdots & & \ddots & \vdots \\ U_{q,1} & \cdots & U_{q,j} & \cdots & U_{q,D} \end{bmatrix} \tag{B5a}$$

$$U_{i,j} = \begin{cases} 1 & if \quad B_i \leq E_j < B_{i+1} \\ 0 & else \end{cases} \tag{B5b}$$

In general, specifying the desired resolution at some distribution order $q_u$ the distribution of any order $u$ can be obtained starting from the preceding order $u-1$ as:

$$\boldsymbol{P}^{[u]} = \boldsymbol{U}^{[u-1,u]} \cdot \boldsymbol{P}^{[u-1]} . \tag{B6}$$

$$\boldsymbol{U}^{[u-1,u]} = \begin{bmatrix} U_{1,1} & U_{1,2} & \cdots & & U_{1,\,q_{u-1}} \\ U_{2,1} & \ddots & \cdots & \cdots & \vdots \\ \vdots & \vdots & U_{i,j} & & U_{i,\,q_{u-1}} \\ & \vdots & & \ddots & \vdots \\ U_{q_u,1} & \cdots & U_{q_u,j} & \cdots & U_{q_u,\,q_{u-1}} \end{bmatrix} \tag{B7a}$$

$$U_{i,j} = \begin{cases} 1 & if \quad B_i \leq E_j < B_{i+1} \\ 0 & else \end{cases} \tag{B7b}$$

$$\boldsymbol{B}_u = \begin{bmatrix} B_1 \\ B_2 \\ \vdots \\ B_{q_u} \end{bmatrix} \tag{B7c}$$

$$\Delta q_u = \frac{E_{max} - E_{min}}{q_u} \, , \quad B_{u_i} = B_{u_{i-1}} + \Delta q_u \, , \quad B_{u_1} = E_{u_{min}} - \frac{\Delta q_u}{2} \tag{B7d)(B7e)(B7f}$$

## Appendix C. Math formulation for Scale Downgrading

The frequency profile associated to a complex language is a representation of the language. In a language made of $D$ different symbols, this representation uses $D$ values to describe the language. The graphical representation of these values is useful because it permits to observe an abstract depiction. Depending on the level of detail the observer intends to appreciate, the $D$ values may or may not be needed. If for some purpose a rough idea of the profile's shape is sufficient, a smaller number of values can be used. If on the contrary, the observer needs to detail tiny changes in the profile, a higher density of dots will be required to draw these changes of direction. Changing the number of symbols used to describe a system constitutes a change of the scale of observation of the system; thus we refer to the process of reducing the number of values used to draw the frequency profile as *downgrading the language scale*.

Consider de language **B** as a set of *D* different symbols. If language **B** is employed to build a *N* symbol long system description, then language **B** can be specified as the set of *D* symbols $Y_i$ and the probability density function **P(Y_i)** which establishes the relative frequencies of appearance of the symbols $f_i$. Thus

$$\boldsymbol{B} = \{Y_1, \dots, Y_i, \dots, Y_D, \boldsymbol{P}(Y_i)\}, \tag{C1}$$

$$P(Y_i) = \frac{f_i}{N}, \quad 1 \leq i \leq D. \tag{C2}$$

At this point language **B** is presented at scale *D*. To include the observation scale of a language as part of the nomenclature, we propose adding a sub-index to the letter representing the language. Then, language **B** at some scale *S*, where *1≤ S ≤ D*, would be denoted as **B**$_{[S]}$. When the index does not appear, it can be assumed the language is expressed at its original and maximum scale. That is **B** = **B**$_{[D]}$. Downgrading a language from scale *D* to scale *S* can be performed by pre multiplying vector **P** with transformation matrix **G**, as indicated below:

$$\boldsymbol{P}_{[S]} = \boldsymbol{G}_{[S,D]} \cdot \boldsymbol{P}_{[D]}. \tag{C3}$$

$$\boldsymbol{G}_{[S,D]} = \begin{bmatrix} G_{1,1} & G_{1,2} & \cdots & & G_{1,D} \\ G_{2,1} & \ddots & \cdots & \cdots & \vdots \\ \vdots & \vdots & G_{i,j} & & G_{i,D} \\ & \vdots & & \ddots & \vdots \\ G_{S,1} & \cdots & G_{S,j} & \cdots & G_{S,D} \end{bmatrix}, \tag{C4a}$$

$$G_{i,j} = \begin{cases} 1 & \text{if } \log_D(j-1) \leq \log_S i < \log_D j \\ 0 & \text{otherwise} \end{cases}, 1 \leq j \leq D, 1 \leq i \leq S, \tag{C4b}$$

$$j = int\,(S^{\log_D i}). \tag{C4c}$$

This procedure for downgrading the language scale is useful given the frequent requirement of expressing text descriptions at the same scale, which is, using the same number of different symbols.

## Appendix D. Information profiles of some selected music

Beethoven's 9th Symphony 4th Movement

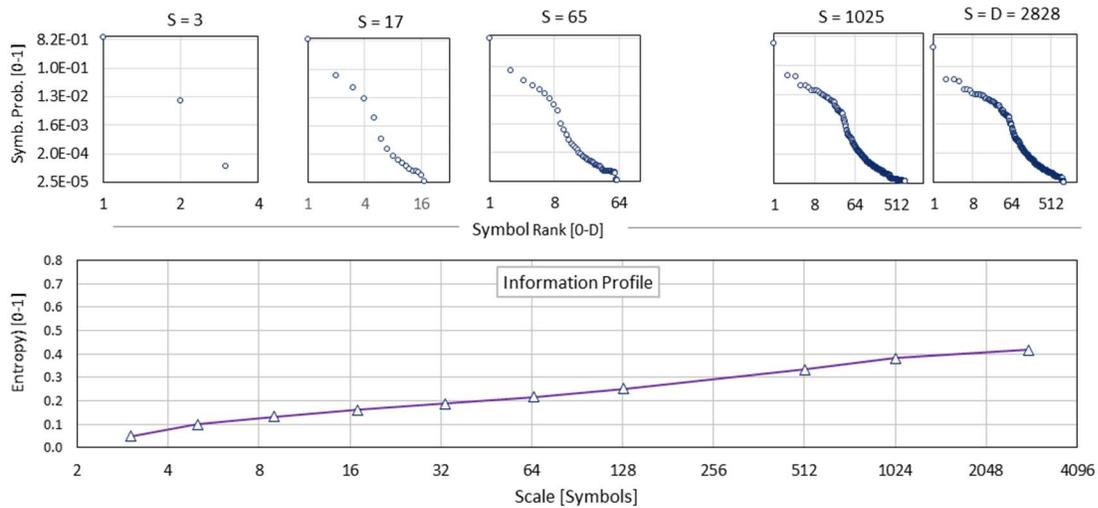

LAURO.Antonio.ValsVenezolanoNro3.Natalia

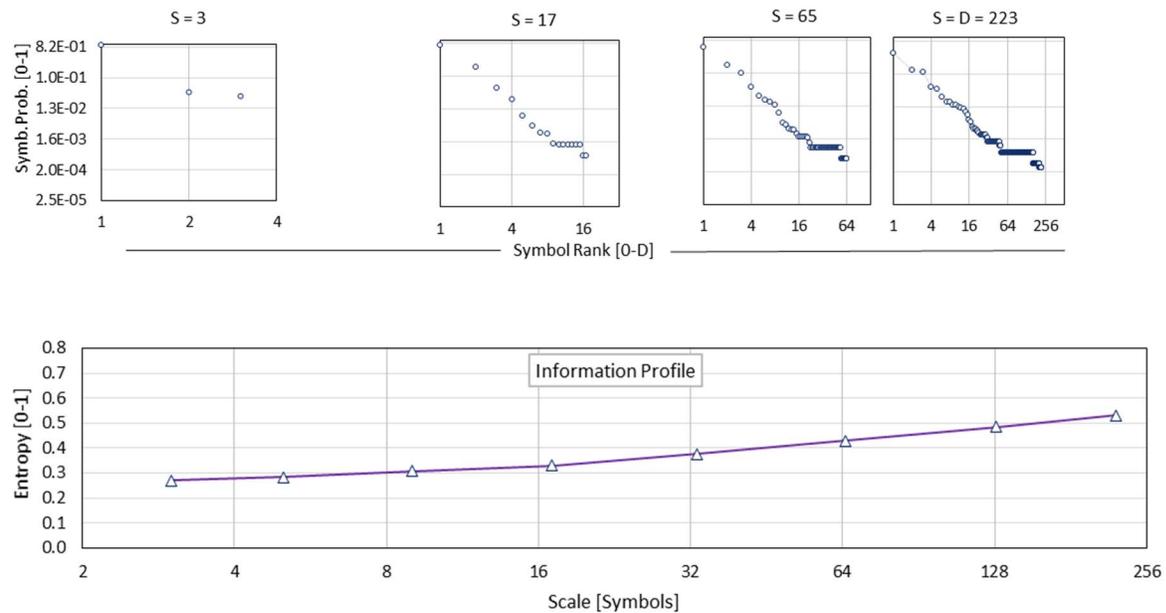

# Appendix E. Music symbol frequency profiles data represented with 128 degrees of freedom

**Symbol probality for several types of music. Observation scale = Diversity = 129 Symbols. From r=1 to r=62**

| rnk. | Medieval | Renais. | Baroque | Classic | Romant. | Impres. | 20th Cty. | Chinese | Raga | M.Thems | Rock | Venez. |
|---|---|---|---|---|---|---|---|---|---|---|---|---|
| 1 | 5.96E-01 | 6.96E-01 | 8.87E-01 | 8.92E-01 | 9.00E-01 | 8.92E-01 | 9.05E-01 | 8.25E-01 | 5.85E-01 | 7.82E-01 | 8.63E-01 | 9.02E-01 |
| 2 | 1.52E-01 | 1.35E-01 | 3.63E-02 | 1.31E-02 | 1.12E-02 | 2.69E-02 | 2.45E-02 | 6.29E-02 | 1.37E-01 | 1.04E-01 | 4.32E-02 | 1.83E-02 |
| 3 | 7.92E-02 | 5.16E-02 | 7.92E-03 | 8.52E-03 | 7.58E-03 | 1.43E-02 | 7.00E-03 | 9.77E-03 | 6.85E-02 | 1.10E-02 | 8.38E-03 | 1.01E-02 |
| 4 | 4.62E-02 | 1.80E-02 | 5.29E-03 | 6.87E-03 | 5.86E-03 | 6.53E-03 | 5.11E-03 | 6.19E-03 | 5.25E-02 | 5.88E-03 | 5.13E-03 | 7.17E-03 |
| 5 | 2.75E-02 | 1.00E-02 | 4.10E-03 | 5.77E-03 | 4.82E-03 | 4.73E-03 | 4.13E-03 | 5.05E-03 | 2.79E-02 | 4.31E-03 | 4.05E-03 | 5.56E-03 |
| 6 | 1.61E-02 | 7.38E-03 | 3.49E-03 | 4.99E-03 | 4.11E-03 | 3.79E-03 | 3.44E-03 | 4.46E-03 | 1.01E-02 | 3.57E-03 | 3.46E-03 | 4.72E-03 |
| 7 | 1.04E-02 | 5.62E-03 | 2.98E-03 | 4.40E-03 | 3.63E-03 | 3.32E-03 | 2.91E-03 | 3.94E-03 | 5.69E-03 | 3.15E-03 | 3.03E-03 | 4.03E-03 |
| 8 | 6.74E-03 | 4.48E-03 | 2.63E-03 | 3.99E-03 | 3.26E-03 | 3.05E-03 | 2.62E-03 | 3.47E-03 | 4.38E-03 | 2.91E-03 | 2.68E-03 | 3.32E-03 |
| 9 | 5.33E-03 | 3.80E-03 | 2.36E-03 | 3.62E-03 | 2.95E-03 | 2.79E-03 | 2.43E-03 | 3.15E-03 | 3.60E-03 | 2.61E-03 | 2.37E-03 | 2.88E-03 |
| 10 | 4.20E-03 | 3.40E-03 | 2.10E-03 | 3.36E-03 | 2.67E-03 | 2.38E-03 | 2.17E-03 | 3.06E-03 | 3.46E-03 | 2.42E-03 | 2.17E-03 | 2.62E-03 |
| 11 | 3.75E-03 | 2.92E-03 | 1.91E-03 | 2.92E-03 | 2.40E-03 | 2.13E-03 | 1.95E-03 | 2.83E-03 | 3.27E-03 | 2.20E-03 | 2.01E-03 | 2.33E-03 |
| 12 | 2.99E-03 | 2.67E-03 | 1.82E-03 | 2.60E-03 | 2.18E-03 | 1.87E-03 | 1.80E-03 | 2.77E-03 | 2.98E-03 | 2.02E-03 | 1.90E-03 | 2.13E-03 |
| 13 | 2.69E-03 | 2.54E-03 | 1.65E-03 | 2.38E-03 | 2.03E-03 | 1.72E-03 | 1.68E-03 | 2.68E-03 | 2.87E-03 | 1.94E-03 | 1.78E-03 | 1.94E-03 |
| 14 | 2.61E-03 | 2.33E-03 | 1.54E-03 | 2.33E-03 | 1.92E-03 | 1.62E-03 | 1.52E-03 | 2.56E-03 | 2.61E-03 | 1.87E-03 | 1.67E-03 | 1.81E-03 |
| 15 | 2.51E-03 | 2.23E-03 | 1.45E-03 | 2.23E-03 | 1.80E-03 | 1.51E-03 | 1.43E-03 | 2.51E-03 | 2.51E-03 | 1.80E-03 | 1.55E-03 | 1.73E-03 |
| 16 | 2.05E-03 | 2.11E-03 | 1.40E-03 | 2.07E-03 | 1.67E-03 | 1.36E-03 | 1.35E-03 | 2.45E-03 | 2.31E-03 | 1.69E-03 | 1.47E-03 | 1.67E-03 |
| 17 | 1.88E-03 | 1.93E-03 | 1.31E-03 | 1.95E-03 | 1.55E-03 | 1.24E-03 | 1.28E-03 | 2.31E-03 | 2.25E-03 | 1.68E-03 | 1.43E-03 | 1.59E-03 |
| 18 | 1.63E-03 | 1.85E-03 | 1.27E-03 | 1.83E-03 | 1.45E-03 | 1.20E-03 | 1.21E-03 | 2.18E-03 | 2.08E-03 | 1.61E-03 | 1.37E-03 | 1.53E-03 |
| 19 | 1.50E-03 | 1.80E-03 | 1.25E-03 | 1.68E-03 | 1.38E-03 | 1.18E-03 | 1.12E-03 | 1.99E-03 | 2.05E-03 | 1.54E-03 | 1.31E-03 | 1.37E-03 |
| 20 | 1.39E-03 | 1.64E-03 | 1.21E-03 | 1.50E-03 | 1.33E-03 | 1.14E-03 | 1.03E-03 | 1.85E-03 | 2.00E-03 | 1.48E-03 | 1.29E-03 | 1.26E-03 |
| 21 | 1.30E-03 | 1.56E-03 | 1.14E-03 | 1.35E-03 | 1.25E-03 | 1.12E-03 | 9.74E-04 | 1.70E-03 | 1.86E-03 | 1.44E-03 | 1.25E-03 | 1.15E-03 |
| 22 | 1.22E-03 | 1.52E-03 | 1.08E-03 | 1.24E-03 | 1.17E-03 | 1.10E-03 | 9.43E-04 | 1.61E-03 | 1.85E-03 | 1.41E-03 | 1.17E-03 | 1.02E-03 |
| 23 | 1.09E-03 | 1.47E-03 | 1.01E-03 | 1.18E-03 | 1.12E-03 | 1.01E-03 | 8.99E-04 | 1.53E-03 | 1.72E-03 | 1.36E-03 | 1.13E-03 | 9.50E-04 |
| 24 | 9.73E-04 | 1.43E-03 | 9.53E-04 | 1.12E-03 | 1.06E-03 | 9.15E-04 | 8.58E-04 | 1.42E-03 | 1.71E-03 | 1.30E-03 | 1.09E-03 | 9.01E-04 |
| 25 | 9.70E-04 | 1.35E-03 | 8.92E-04 | 1.05E-03 | 1.02E-03 | 8.50E-04 | 8.26E-04 | 1.31E-03 | 1.62E-03 | 1.23E-03 | 1.07E-03 | 8.80E-04 |
| 26 | 8.85E-04 | 1.27E-03 | 8.40E-04 | 9.83E-04 | 9.75E-04 | 7.91E-04 | 7.94E-04 | 1.28E-03 | 1.62E-03 | 1.18E-03 | 1.05E-03 | 8.60E-04 |
| 27 | 8.29E-04 | 1.20E-03 | 7.95E-04 | 9.39E-04 | 9.31E-04 | 7.71E-04 | 7.52E-04 | 1.26E-03 | 1.54E-03 | 1.14E-03 | 1.01E-03 | 8.39E-04 |
| 28 | 7.93E-04 | 1.14E-03 | 7.51E-04 | 9.12E-04 | 8.95E-04 | 7.29E-04 | 7.24E-04 | 1.24E-03 | 1.53E-03 | 1.10E-03 | 9.92E-04 | 7.83E-04 |
| 29 | 7.53E-04 | 1.11E-03 | 7.09E-04 | 8.85E-04 | 8.55E-04 | 6.63E-04 | 6.96E-04 | 1.20E-03 | 1.45E-03 | 1.07E-03 | 9.42E-04 | 7.02E-04 |
| 30 | 7.04E-04 | 1.04E-03 | 6.70E-04 | 8.47E-04 | 8.21E-04 | 6.28E-04 | 6.53E-04 | 1.11E-03 | 1.44E-03 | 1.04E-03 | 9.16E-04 | 6.45E-04 |
| 31 | 6.65E-04 | 1.02E-03 | 6.44E-04 | 8.08E-04 | 7.93E-04 | 6.08E-04 | 6.20E-04 | 1.06E-03 | 1.39E-03 | 1.02E-03 | 9.01E-04 | 6.21E-04 |
| 32 | 6.32E-04 | 1.00E-03 | 6.23E-04 | 7.51E-04 | 7.69E-04 | 5.95E-04 | 5.89E-04 | 1.00E-03 | 1.36E-03 | 1.01E-03 | 8.79E-04 | 5.79E-04 |
| 33 | 6.07E-04 | 9.96E-04 | 6.11E-04 | 6.93E-04 | 7.41E-04 | 5.72E-04 | 5.55E-04 | 9.42E-04 | 1.32E-03 | 1.00E-03 | 8.49E-04 | 5.38E-04 |
| 34 | 5.89E-04 | 9.78E-04 | 5.89E-04 | 6.42E-04 | 7.09E-04 | 5.43E-04 | 5.18E-04 | 8.78E-04 | 1.32E-03 | 9.78E-04 | 7.99E-04 | 5.15E-04 |
| 35 | 5.70E-04 | 9.31E-04 | 5.65E-04 | 6.23E-04 | 6.85E-04 | 5.08E-04 | 4.98E-04 | 8.47E-04 | 1.32E-03 | 9.61E-04 | 7.79E-04 | 4.85E-04 |
| 36 | 5.30E-04 | 8.97E-04 | 5.47E-04 | 6.16E-04 | 6.54E-04 | 4.75E-04 | 4.82E-04 | 8.34E-04 | 1.25E-03 | 9.61E-04 | 7.69E-04 | 4.51E-04 |
| 37 | 5.14E-04 | 8.23E-04 | 5.33E-04 | 6.08E-04 | 6.22E-04 | 4.46E-04 | 4.67E-04 | 8.28E-04 | 1.22E-03 | 9.44E-04 | 7.51E-04 | 4.27E-04 |
| 38 | 4.86E-04 | 7.82E-04 | 5.15E-04 | 5.93E-04 | 5.95E-04 | 4.25E-04 | 4.61E-04 | 8.10E-04 | 1.15E-03 | 9.22E-04 | 7.34E-04 | 4.12E-04 |
| 39 | 4.74E-04 | 7.46E-04 | 5.01E-04 | 5.76E-04 | 5.65E-04 | 4.09E-04 | 4.46E-04 | 7.88E-04 | 1.15E-03 | 9.04E-04 | 7.28E-04 | 3.98E-04 |
| 40 | 4.65E-04 | 7.04E-04 | 4.77E-04 | 5.56E-04 | 5.44E-04 | 3.93E-04 | 4.31E-04 | 7.81E-04 | 1.12E-03 | 8.79E-04 | 7.22E-04 | 3.79E-04 |
| 41 | 4.65E-04 | 6.85E-04 | 4.67E-04 | 5.19E-04 | 5.25E-04 | 3.78E-04 | 4.22E-04 | 7.70E-04 | 1.11E-03 | 8.60E-04 | 7.14E-04 | 3.50E-04 |
| 42 | 4.63E-04 | 6.79E-04 | 4.58E-04 | 4.81E-04 | 5.16E-04 | 3.64E-04 | 4.14E-04 | 7.32E-04 | 1.08E-03 | 8.40E-04 | 7.10E-04 | 3.22E-04 |
| 43 | 4.44E-04 | 6.78E-04 | 4.48E-04 | 4.59E-04 | 5.02E-04 | 3.47E-04 | 4.02E-04 | 6.95E-04 | 1.05E-03 | 8.25E-04 | 7.04E-04 | 3.07E-04 |
| 44 | 4.35E-04 | 6.36E-04 | 4.32E-04 | 4.42E-04 | 4.83E-04 | 3.28E-04 | 3.86E-04 | 6.59E-04 | 1.04E-03 | 7.99E-04 | 6.94E-04 | 2.97E-04 |
| 45 | 4.26E-04 | 6.27E-04 | 4.25E-04 | 4.23E-04 | 4.70E-04 | 3.15E-04 | 3.77E-04 | 6.28E-04 | 1.01E-03 | 7.91E-04 | 6.77E-04 | 2.82E-04 |
| 46 | 4.18E-04 | 5.99E-04 | 4.16E-04 | 4.06E-04 | 4.58E-04 | 3.08E-04 | 3.68E-04 | 5.93E-04 | 9.88E-04 | 7.79E-04 | 6.55E-04 | 2.68E-04 |
| 47 | 3.93E-04 | 5.74E-04 | 4.07E-04 | 3.83E-04 | 4.47E-04 | 3.01E-04 | 3.55E-04 | 5.76E-04 | 9.21E-04 | 7.55E-04 | 6.39E-04 | 2.60E-04 |
| 48 | 3.93E-04 | 5.44E-04 | 3.97E-04 | 3.65E-04 | 4.33E-04 | 2.93E-04 | 3.37E-04 | 5.63E-04 | 8.91E-04 | 7.31E-04 | 6.24E-04 | 2.51E-04 |
| 49 | 3.89E-04 | 5.18E-04 | 3.92E-04 | 3.49E-04 | 4.21E-04 | 2.85E-04 | 3.28E-04 | 5.40E-04 | 8.79E-04 | 7.21E-04 | 6.05E-04 | 2.40E-04 |
| 50 | 3.84E-04 | 5.04E-04 | 3.85E-04 | 3.30E-04 | 4.10E-04 | 2.76E-04 | 3.21E-04 | 5.22E-04 | 8.65E-04 | 7.14E-04 | 5.80E-04 | 2.23E-04 |
| 51 | 3.66E-04 | 4.91E-04 | 3.79E-04 | 3.16E-04 | 4.03E-04 | 2.64E-04 | 3.07E-04 | 5.21E-04 | 8.59E-04 | 7.12E-04 | 5.53E-04 | 2.13E-04 |
| 52 | 3.61E-04 | 4.63E-04 | 3.71E-04 | 3.10E-04 | 3.98E-04 | 2.50E-04 | 2.99E-04 | 5.14E-04 | 8.46E-04 | 6.94E-04 | 5.38E-04 | 2.08E-04 |
| 53 | 3.58E-04 | 4.46E-04 | 3.59E-04 | 3.05E-04 | 3.90E-04 | 2.42E-04 | 2.91E-04 | 5.09E-04 | 8.41E-04 | 6.76E-04 | 5.21E-04 | 2.02E-04 |
| 54 | 3.51E-04 | 4.28E-04 | 3.47E-04 | 2.99E-04 | 3.82E-04 | 2.36E-04 | 2.87E-04 | 5.00E-04 | 8.07E-04 | 6.53E-04 | 5.07E-04 | 1.99E-04 |
| 55 | 3.32E-04 | 3.97E-04 | 3.32E-04 | 2.94E-04 | 3.73E-04 | 2.30E-04 | 2.81E-04 | 4.88E-04 | 8.02E-04 | 6.44E-04 | 4.97E-04 | 1.94E-04 |
| 56 | 3.20E-04 | 3.76E-04 | 3.25E-04 | 2.87E-04 | 3.62E-04 | 2.24E-04 | 2.75E-04 | 4.83E-04 | 7.97E-04 | 6.24E-04 | 4.81E-04 | 1.88E-04 |
| 57 | 3.16E-04 | 3.56E-04 | 3.16E-04 | 2.81E-04 | 3.52E-04 | 2.16E-04 | 2.66E-04 | 4.65E-04 | 7.89E-04 | 6.05E-04 | 4.64E-04 | 1.75E-04 |
| 58 | 3.07E-04 | 3.54E-04 | 3.12E-04 | 2.74E-04 | 3.43E-04 | 2.09E-04 | 2.59E-04 | 4.58E-04 | 7.89E-04 | 5.93E-04 | 4.55E-04 | 1.67E-04 |
| 59 | 2.96E-04 | 3.41E-04 | 3.06E-04 | 2.63E-04 | 3.36E-04 | 2.03E-04 | 2.52E-04 | 4.43E-04 | 7.74E-04 | 5.76E-04 | 4.45E-04 | 1.62E-04 |
| 60 | 2.95E-04 | 3.39E-04 | 3.01E-04 | 2.51E-04 | 3.23E-04 | 1.96E-04 | 2.46E-04 | 4.38E-04 | 7.70E-04 | 5.70E-04 | 4.37E-04 | 1.54E-04 |
| 61 | 2.93E-04 | 3.30E-04 | 2.92E-04 | 2.45E-04 | 3.13E-04 | 1.89E-04 | 2.45E-04 | 4.30E-04 | 7.68E-04 | 5.67E-04 | 4.28E-04 | 1.49E-04 |
| 62 | 2.87E-04 | 3.22E-04 | 2.85E-04 | 2.39E-04 | 3.04E-04 | 1.83E-04 | 2.34E-04 | 4.20E-04 | 7.65E-04 | 5.53E-04 | 4.22E-04 | 1.40E-04 |

**Symbol probality for several types of music. Observation scale = Diversity = 129 Symbols. From r=63 to r=129**

| rnk. | Medieval | Renais. | Baroque | Classic | Romant. | Impres. | 20th Cty. | Chinese | Raga | M.Thems | Rock | Venez. |
|---|---|---|---|---|---|---|---|---|---|---|---|---|
| 63 | 2.87E-04 | 3.16E-04 | 2.80E-04 | 2.35E-04 | 2.94E-04 | 1.78E-04 | 2.21E-04 | 3.99E-04 | 7.52E-04 | 5.43E-04 | 4.16E-04 | 1.36E-04 |
| 64 | 2.87E-04 | 3.05E-04 | 2.71E-04 | 2.31E-04 | 2.85E-04 | 1.73E-04 | 2.15E-04 | 3.73E-04 | 7.46E-04 | 5.24E-04 | 4.13E-04 | 1.29E-04 |
| 65 | 2.86E-04 | 2.98E-04 | 2.62E-04 | 2.24E-04 | 2.75E-04 | 1.66E-04 | 2.09E-04 | 3.53E-04 | 7.45E-04 | 5.19E-04 | 4.04E-04 | 1.21E-04 |
| 66 | 2.83E-04 | 2.96E-04 | 2.60E-04 | 2.15E-04 | 2.67E-04 | 1.60E-04 | 2.04E-04 | 3.39E-04 | 7.44E-04 | 5.15E-04 | 3.90E-04 | 1.13E-04 |
| 67 | 2.82E-04 | 2.93E-04 | 2.55E-04 | 2.08E-04 | 2.61E-04 | 1.56E-04 | 2.00E-04 | 3.29E-04 | 7.23E-04 | 5.07E-04 | 3.85E-04 | 1.12E-04 |
| 68 | 2.81E-04 | 2.82E-04 | 2.47E-04 | 2.03E-04 | 2.58E-04 | 1.53E-04 | 1.93E-04 | 3.29E-04 | 7.23E-04 | 4.98E-04 | 3.78E-04 | 1.11E-04 |
| 69 | 2.70E-04 | 2.74E-04 | 2.41E-04 | 1.96E-04 | 2.54E-04 | 1.51E-04 | 1.88E-04 | 3.22E-04 | 6.98E-04 | 4.87E-04 | 3.70E-04 | 1.10E-04 |
| 70 | 2.58E-04 | 2.62E-04 | 2.35E-04 | 1.91E-04 | 2.49E-04 | 1.49E-04 | 1.80E-04 | 3.10E-04 | 6.97E-04 | 4.83E-04 | 3.68E-04 | 1.08E-04 |
| 71 | 2.43E-04 | 2.54E-04 | 2.30E-04 | 1.88E-04 | 2.45E-04 | 1.48E-04 | 1.70E-04 | 3.05E-04 | 6.96E-04 | 4.82E-04 | 3.67E-04 | 1.07E-04 |
| 72 | 2.41E-04 | 2.53E-04 | 2.28E-04 | 1.84E-04 | 2.39E-04 | 1.47E-04 | 1.63E-04 | 3.02E-04 | 6.96E-04 | 4.79E-04 | 3.64E-04 | 1.05E-04 |
| 73 | 2.40E-04 | 2.41E-04 | 2.20E-04 | 1.81E-04 | 2.35E-04 | 1.45E-04 | 1.56E-04 | 3.02E-04 | 6.95E-04 | 4.76E-04 | 3.60E-04 | 1.03E-04 |
| 74 | 2.34E-04 | 2.29E-04 | 2.14E-04 | 1.76E-04 | 2.30E-04 | 1.44E-04 | 1.51E-04 | 2.98E-04 | 6.76E-04 | 4.73E-04 | 3.55E-04 | 1.02E-04 |
| 75 | 2.27E-04 | 2.23E-04 | 2.09E-04 | 1.70E-04 | 2.24E-04 | 1.41E-04 | 1.48E-04 | 2.89E-04 | 6.75E-04 | 4.69E-04 | 3.46E-04 | 9.84E-05 |
| 76 | 2.19E-04 | 2.08E-04 | 2.06E-04 | 1.62E-04 | 2.19E-04 | 1.40E-04 | 1.46E-04 | 2.88E-04 | 6.62E-04 | 4.61E-04 | 3.39E-04 | 9.51E-05 |
| 77 | 2.10E-04 | 1.99E-04 | 2.03E-04 | 1.58E-04 | 2.15E-04 | 1.39E-04 | 1.46E-04 | 2.88E-04 | 6.55E-04 | 4.50E-04 | 3.34E-04 | 9.19E-05 |
| 78 | 1.99E-04 | 1.95E-04 | 2.00E-04 | 1.53E-04 | 2.12E-04 | 1.36E-04 | 1.45E-04 | 2.85E-04 | 6.53E-04 | 4.44E-04 | 3.32E-04 | 8.92E-05 |
| 79 | 1.95E-04 | 1.87E-04 | 1.98E-04 | 1.50E-04 | 2.09E-04 | 1.31E-04 | 1.44E-04 | 2.84E-04 | 6.34E-04 | 4.44E-04 | 3.28E-04 | 8.35E-05 |
| 80 | 1.91E-04 | 1.87E-04 | 1.93E-04 | 1.46E-04 | 2.07E-04 | 1.26E-04 | 1.43E-04 | 2.81E-04 | 6.28E-04 | 4.36E-04 | 3.23E-04 | 7.84E-05 |
| 81 | 1.89E-04 | 1.85E-04 | 1.91E-04 | 1.42E-04 | 2.04E-04 | 1.21E-04 | 1.42E-04 | 2.80E-04 | 5.85E-04 | 4.33E-04 | 3.21E-04 | 7.48E-05 |
| 82 | 1.86E-04 | 1.80E-04 | 1.85E-04 | 1.38E-04 | 2.01E-04 | 1.16E-04 | 1.41E-04 | 2.79E-04 | 5.74E-04 | 4.31E-04 | 3.14E-04 | 7.09E-05 |
| 83 | 1.80E-04 | 1.78E-04 | 1.77E-04 | 1.35E-04 | 1.97E-04 | 1.14E-04 | 1.40E-04 | 2.77E-04 | 5.61E-04 | 4.27E-04 | 3.11E-04 | 6.56E-05 |
| 84 | 1.77E-04 | 1.77E-04 | 1.69E-04 | 1.32E-04 | 1.93E-04 | 1.13E-04 | 1.37E-04 | 2.66E-04 | 5.56E-04 | 4.23E-04 | 3.09E-04 | 6.37E-05 |
| 85 | 1.76E-04 | 1.76E-04 | 1.64E-04 | 1.30E-04 | 1.88E-04 | 1.08E-04 | 1.35E-04 | 2.58E-04 | 5.44E-04 | 4.18E-04 | 3.04E-04 | 5.98E-05 |
| 86 | 1.74E-04 | 1.76E-04 | 1.61E-04 | 1.27E-04 | 1.84E-04 | 1.05E-04 | 1.33E-04 | 2.58E-04 | 5.44E-04 | 4.13E-04 | 2.98E-04 | 5.72E-05 |
| 87 | 1.69E-04 | 1.70E-04 | 1.59E-04 | 1.25E-04 | 1.80E-04 | 1.02E-04 | 1.29E-04 | 2.55E-04 | 4.90E-04 | 4.04E-04 | 2.97E-04 | 5.63E-05 |
| 88 | 1.65E-04 | 1.69E-04 | 1.56E-04 | 1.23E-04 | 1.77E-04 | 1.01E-04 | 1.28E-04 | 2.48E-04 | 4.88E-04 | 4.02E-04 | 2.93E-04 | 5.56E-05 |
| 89 | 1.63E-04 | 1.68E-04 | 1.55E-04 | 1.20E-04 | 1.74E-04 | 9.96E-05 | 1.26E-04 | 2.46E-04 | 4.79E-04 | 3.96E-04 | 2.78E-04 | 5.45E-05 |
| 90 | 1.62E-04 | 1.66E-04 | 1.52E-04 | 1.19E-04 | 1.72E-04 | 9.72E-05 | 1.24E-04 | 2.39E-04 | 4.71E-04 | 3.89E-04 | 2.66E-04 | 5.36E-05 |
| 91 | 1.59E-04 | 1.65E-04 | 1.52E-04 | 1.17E-04 | 1.69E-04 | 9.48E-05 | 1.21E-04 | 2.29E-04 | 4.68E-04 | 3.88E-04 | 2.62E-04 | 5.28E-05 |
| 92 | 1.59E-04 | 1.62E-04 | 1.50E-04 | 1.14E-04 | 1.67E-04 | 9.20E-05 | 1.11E-04 | 2.23E-04 | 4.31E-04 | 3.87E-04 | 2.59E-04 | 5.22E-05 |
| 93 | 1.58E-04 | 1.62E-04 | 1.46E-04 | 1.08E-04 | 1.62E-04 | 8.61E-05 | 1.02E-04 | 2.22E-04 | 4.27E-04 | 3.78E-04 | 2.57E-04 | 5.17E-05 |
| 94 | 1.49E-04 | 1.61E-04 | 1.42E-04 | 1.04E-04 | 1.59E-04 | 8.10E-05 | 9.62E-05 | 2.21E-04 | 4.02E-04 | 3.71E-04 | 2.55E-04 | 5.06E-05 |
| 95 | 1.49E-04 | 1.60E-04 | 1.40E-04 | 1.02E-04 | 1.54E-04 | 7.63E-05 | 8.45E-05 | 2.20E-04 | 4.00E-04 | 3.64E-04 | 2.53E-04 | 4.97E-05 |
| 96 | 1.47E-04 | 1.60E-04 | 1.34E-04 | 9.98E-05 | 1.50E-04 | 7.40E-05 | 7.95E-05 | 2.18E-04 | 3.97E-04 | 3.63E-04 | 2.51E-04 | 4.93E-05 |
| 97 | 1.47E-04 | 1.57E-04 | 1.29E-04 | 9.79E-05 | 1.48E-04 | 7.25E-05 | 7.68E-05 | 2.18E-04 | 3.95E-04 | 3.47E-04 | 2.48E-04 | 4.88E-05 |
| 98 | 1.47E-04 | 1.50E-04 | 1.21E-04 | 9.55E-05 | 1.45E-04 | 7.04E-05 | 7.57E-05 | 2.17E-04 | 3.91E-04 | 3.40E-04 | 2.40E-04 | 4.85E-05 |
| 99 | 1.44E-04 | 1.48E-04 | 1.14E-04 | 9.33E-05 | 1.41E-04 | 6.87E-05 | 7.43E-05 | 2.17E-04 | 3.89E-04 | 3.34E-04 | 2.30E-04 | 4.72E-05 |
| 100 | 1.42E-04 | 1.45E-04 | 1.04E-04 | 8.67E-05 | 1.34E-04 | 6.61E-05 | 7.28E-05 | 2.13E-04 | 3.85E-04 | 3.28E-04 | 2.21E-04 | 4.59E-05 |
| 101 | 1.42E-04 | 1.39E-04 | 1.00E-04 | 7.93E-05 | 1.28E-04 | 6.30E-05 | 7.15E-05 | 2.13E-04 | 3.81E-04 | 3.28E-04 | 2.16E-04 | 4.05E-05 |
| 102 | 1.41E-04 | 1.31E-04 | 9.44E-05 | 7.56E-05 | 1.25E-04 | 5.92E-05 | 6.90E-05 | 2.08E-04 | 3.76E-04 | 3.18E-04 | 2.11E-04 | 3.64E-05 |
| 103 | 1.30E-04 | 1.26E-04 | 8.71E-05 | 7.37E-05 | 1.22E-04 | 5.60E-05 | 6.51E-05 | 1.85E-04 | 3.62E-04 | 3.15E-04 | 2.06E-04 | 3.44E-05 |
| 104 | 1.26E-04 | 1.17E-04 | 8.18E-05 | 7.23E-05 | 1.19E-04 | 5.27E-05 | 6.11E-05 | 1.60E-04 | 3.55E-04 | 3.10E-04 | 2.01E-04 | 3.34E-05 |
| 105 | 1.25E-04 | 1.10E-04 | 7.87E-05 | 7.06E-05 | 1.16E-04 | 5.08E-05 | 5.76E-05 | 1.56E-04 | 3.34E-04 | 3.05E-04 | 1.97E-04 | 3.24E-05 |
| 106 | 1.22E-04 | 1.02E-04 | 7.67E-05 | 6.92E-05 | 1.11E-04 | 4.89E-05 | 5.28E-05 | 1.54E-04 | 3.28E-04 | 3.04E-04 | 1.91E-04 | 3.15E-05 |
| 107 | 1.20E-04 | 9.69E-05 | 7.47E-05 | 6.85E-05 | 1.06E-04 | 4.57E-05 | 5.16E-05 | 1.53E-04 | 3.14E-04 | 2.95E-04 | 1.86E-04 | 3.13E-05 |
| 108 | 1.19E-04 | 9.34E-05 | 7.18E-05 | 6.74E-05 | 1.01E-04 | 4.14E-05 | 5.12E-05 | 1.51E-04 | 3.12E-04 | 2.94E-04 | 1.83E-04 | 3.08E-05 |
| 109 | 1.16E-04 | 9.16E-05 | 6.78E-05 | 6.59E-05 | 9.85E-05 | 3.85E-05 | 5.09E-05 | 1.50E-04 | 2.84E-04 | 2.90E-04 | 1.77E-04 | 2.84E-05 |
| 110 | 1.14E-04 | 8.90E-05 | 6.10E-05 | 6.52E-05 | 9.49E-05 | 3.54E-05 | 5.01E-05 | 1.50E-04 | 2.62E-04 | 2.89E-04 | 1.70E-04 | 2.54E-05 |
| 111 | 1.11E-04 | 8.67E-05 | 5.65E-05 | 6.39E-05 | 9.16E-05 | 3.43E-05 | 4.81E-05 | 1.49E-04 | 2.31E-04 | 2.88E-04 | 1.67E-04 | 2.44E-05 |
| 112 | 1.09E-04 | 8.48E-05 | 5.22E-05 | 6.29E-05 | 8.85E-05 | 3.39E-05 | 4.43E-05 | 1.49E-04 | 2.19E-04 | 2.87E-04 | 1.64E-04 | 2.41E-05 |
| 113 | 1.06E-04 | 8.14E-05 | 4.75E-05 | 6.22E-05 | 8.62E-05 | 3.35E-05 | 4.25E-05 | 1.47E-04 | 2.19E-04 | 2.86E-04 | 1.58E-04 | 2.36E-05 |
| 114 | 1.02E-04 | 7.70E-05 | 4.43E-05 | 6.11E-05 | 8.49E-05 | 3.29E-05 | 4.13E-05 | 1.47E-04 | 2.13E-04 | 2.85E-04 | 1.51E-04 | 1.91E-05 |
| 115 | 1.01E-04 | 7.28E-05 | 4.21E-05 | 5.90E-05 | 8.32E-05 | 3.24E-05 | 4.04E-05 | 1.46E-04 | 2.13E-04 | 2.85E-04 | 1.47E-04 | 1.65E-05 |
| 116 | 9.85E-05 | 6.97E-05 | 4.06E-05 | 5.32E-05 | 8.12E-05 | 3.19E-05 | 4.00E-05 | 1.46E-04 | 2.13E-04 | 2.82E-04 | 1.45E-04 | 1.64E-05 |
| 117 | 9.72E-05 | 6.90E-05 | 3.91E-05 | 4.64E-05 | 7.88E-05 | 3.13E-05 | 3.96E-05 | 1.44E-04 | 2.13E-04 | 2.72E-04 | 1.42E-04 | 1.62E-05 |
| 118 | 9.09E-05 | 6.81E-05 | 3.82E-05 | 4.25E-05 | 7.53E-05 | 3.07E-05 | 3.92E-05 | 1.43E-04 | 2.10E-04 | 2.60E-04 | 1.37E-04 | 1.61E-05 |
| 119 | 8.41E-05 | 6.61E-05 | 3.62E-05 | 4.12E-05 | 6.98E-05 | 3.02E-05 | 3.87E-05 | 1.43E-04 | 2.10E-04 | 2.57E-04 | 1.30E-04 | 1.61E-05 |
| 120 | 8.12E-05 | 6.49E-05 | 3.18E-05 | 4.05E-05 | 6.55E-05 | 2.93E-05 | 3.85E-05 | 1.42E-04 | 2.07E-04 | 2.51E-04 | 1.28E-04 | 1.60E-05 |
| 121 | 7.91E-05 | 6.36E-05 | 3.00E-05 | 4.00E-05 | 6.24E-05 | 2.84E-05 | 3.84E-05 | 1.40E-04 | 1.99E-04 | 2.39E-04 | 1.26E-04 | 1.60E-05 |
| 122 | 7.80E-05 | 6.01E-05 | 2.90E-05 | 3.93E-05 | 6.03E-05 | 2.77E-05 | 3.80E-05 | 1.39E-04 | 1.98E-04 | 2.24E-04 | 1.23E-04 | 1.58E-05 |
| 123 | 7.28E-05 | 5.95E-05 | 2.82E-05 | 3.84E-05 | 5.81E-05 | 2.71E-05 | 3.76E-05 | 1.30E-04 | 1.94E-04 | 2.09E-04 | 1.17E-04 | 1.57E-05 |
| 124 | 6.71E-05 | 5.88E-05 | 2.75E-05 | 3.77E-05 | 5.33E-05 | 2.69E-05 | 3.42E-05 | 1.11E-04 | 1.94E-04 | 1.96E-04 | 1.06E-04 | 1.56E-05 |
| 125 | 6.19E-05 | 5.79E-05 | 2.61E-05 | 3.68E-05 | 4.43E-05 | 2.67E-05 | 2.49E-05 | 1.07E-04 | 1.14E-04 | 1.77E-04 | 9.94E-05 | 1.55E-05 |
| 126 | 5.86E-05 | 5.71E-05 | 2.15E-05 | 3.50E-05 | 4.12E-05 | 2.63E-05 | 2.12E-05 | 8.77E-05 | 1.10E-04 | 1.60E-04 | 9.19E-05 | 1.28E-05 |
| 127 | 5.30E-05 | 5.17E-05 | 1.90E-05 | 3.25E-05 | 3.90E-05 | 2.39E-05 | 2.00E-05 | 7.54E-05 | 1.06E-04 | 1.51E-04 | 8.17E-05 | 8.12E-06 |
| 128 | 4.81E-05 | 3.54E-05 | 1.52E-05 | 2.26E-05 | 3.52E-05 | 1.60E-05 | 1.94E-05 | 7.36E-05 | 1.05E-04 | 1.44E-04 | 7.15E-05 | 8.02E-06 |
| 129 | 3.75E-05 | 3.08E-05 | 1.25E-05 | 1.95E-05 | 2.53E-05 | 1.36E-05 | 1.90E-05 | 7.15E-05 | 9.99E-05 | 1.37E-04 | 5.91E-05 | 7.88E-06 |

## Appendix F. Symbol Ranked Frequency profiles for 12 different types of music

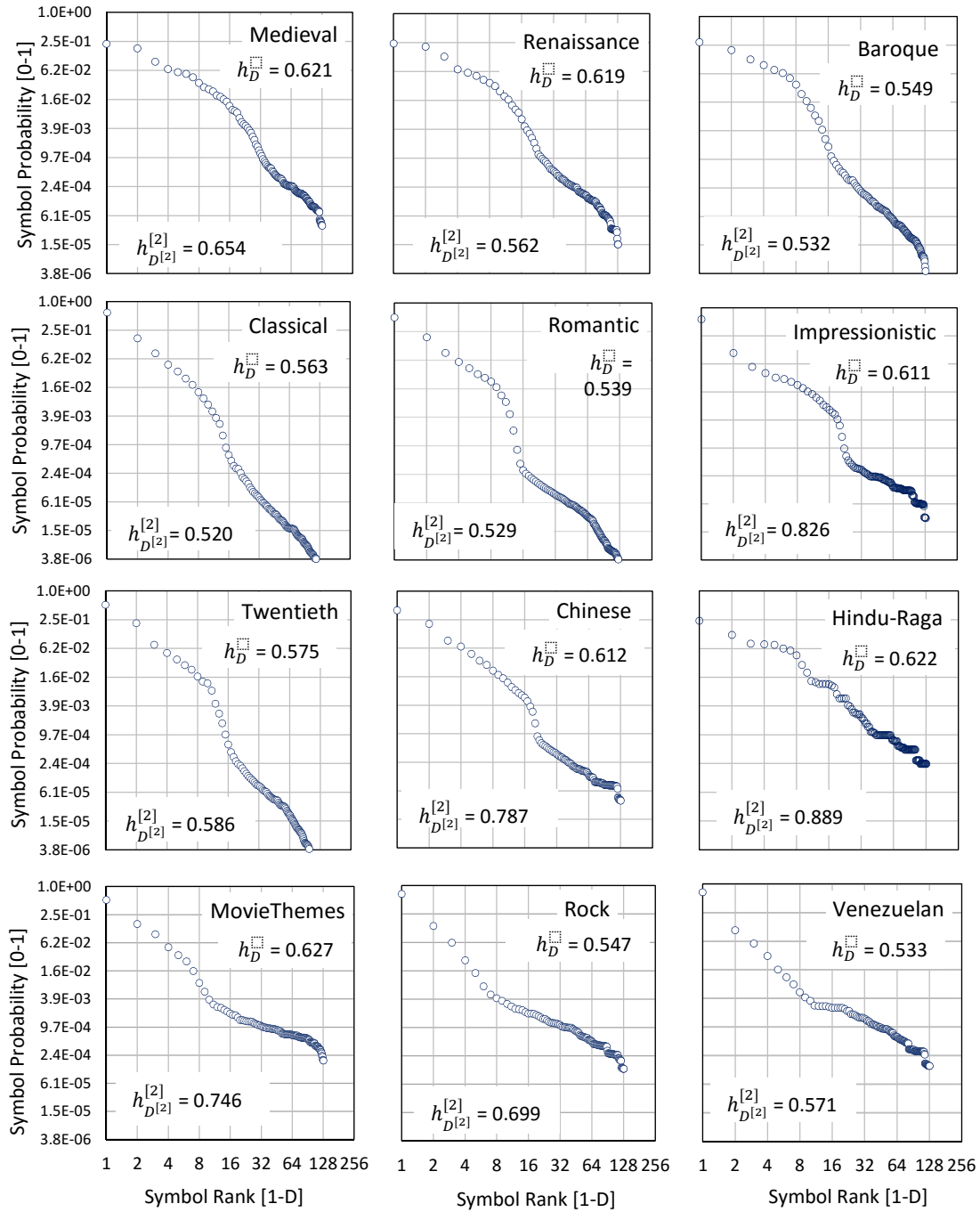

# Appendix G. 2nd order symbol frequency profiles for 12 different types of music

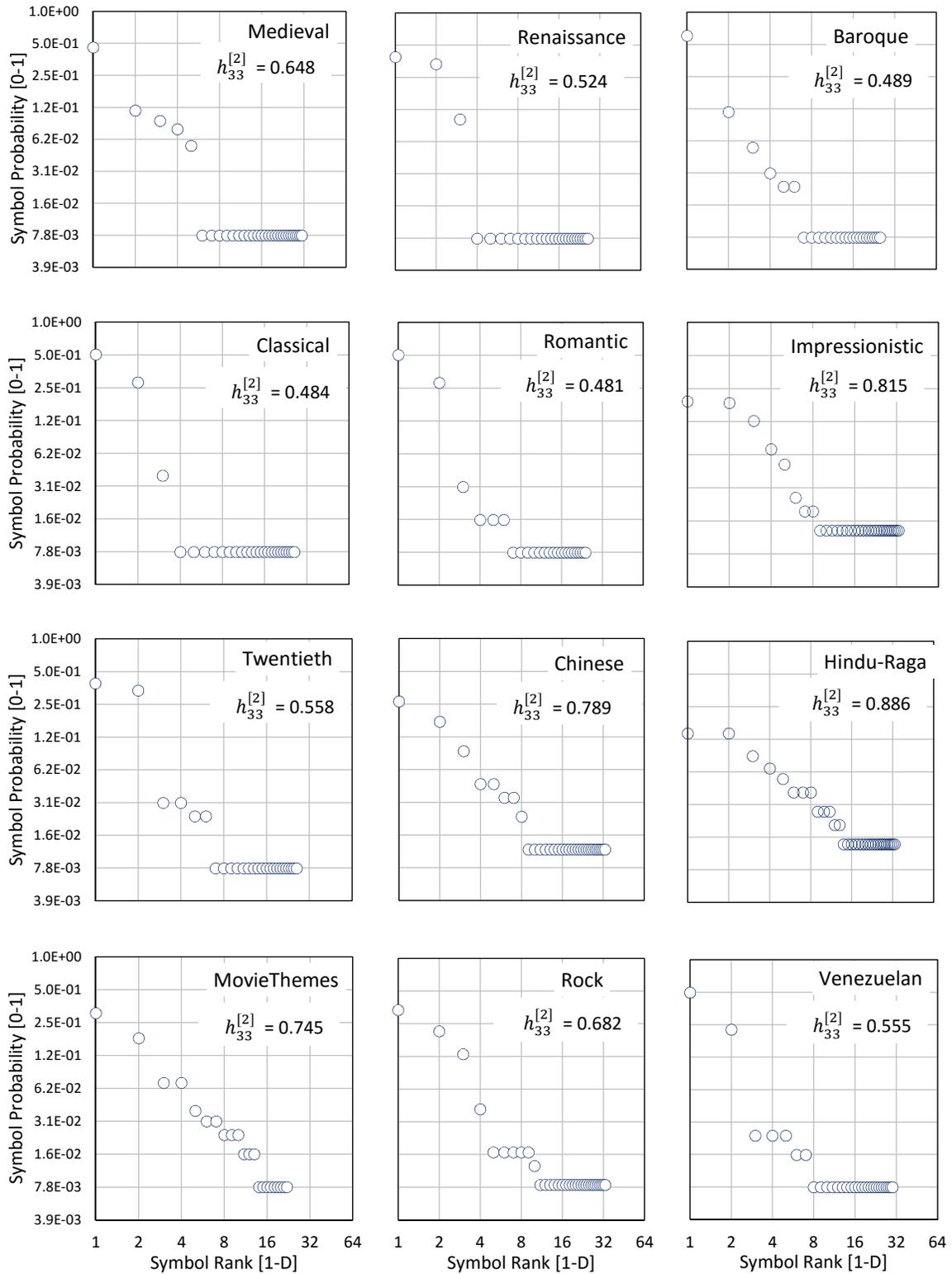

## Appendix H. Music profiles for several composers

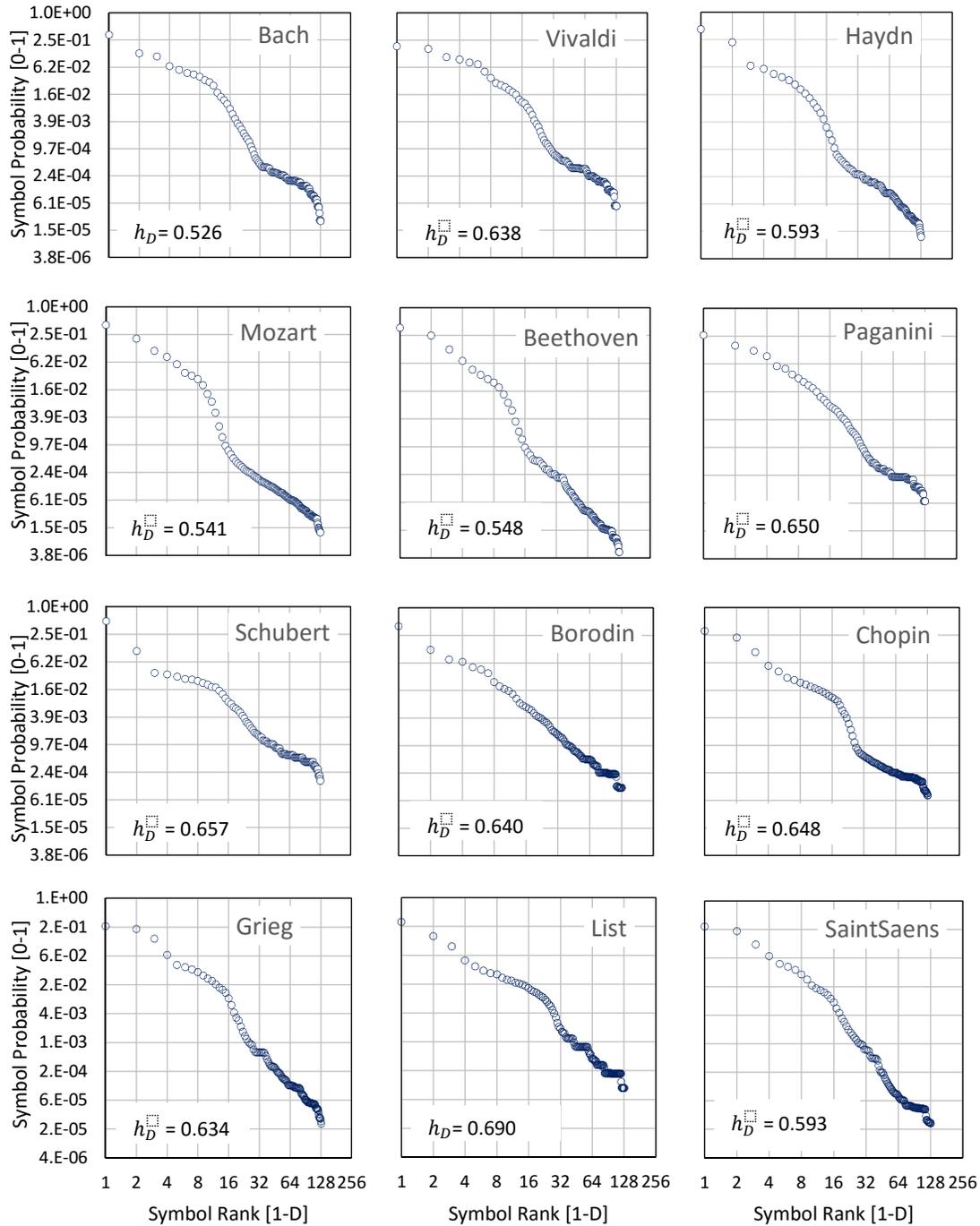

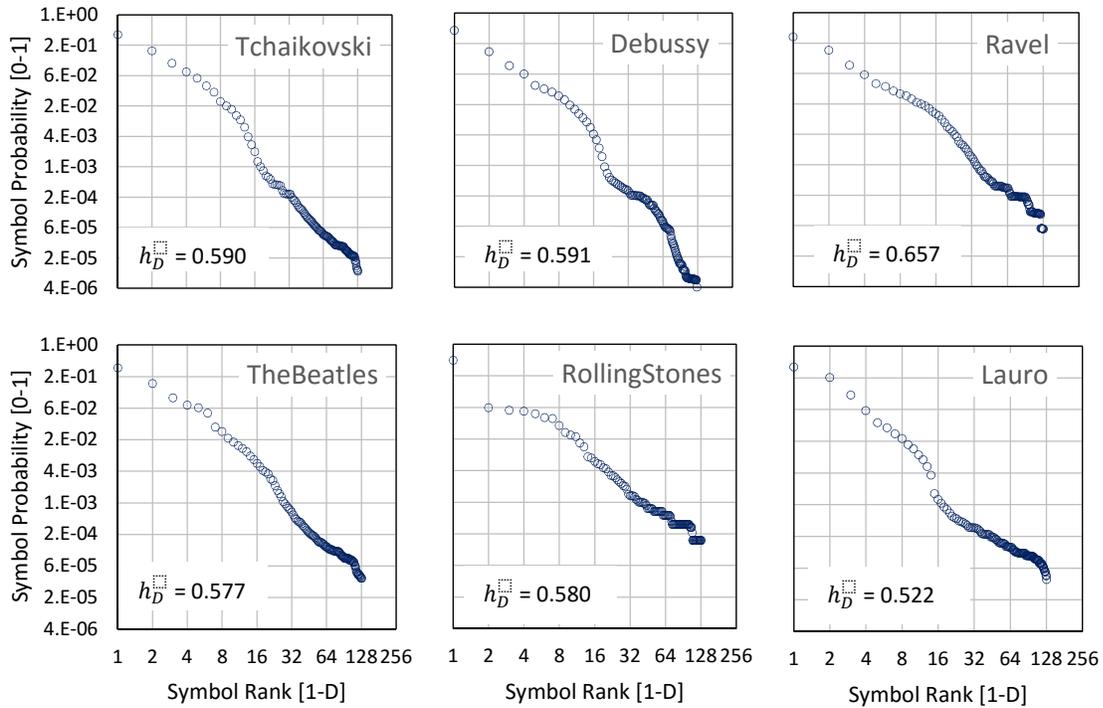

**Appendix I. Change of Specific diversity and entropy over time for different types of music.**

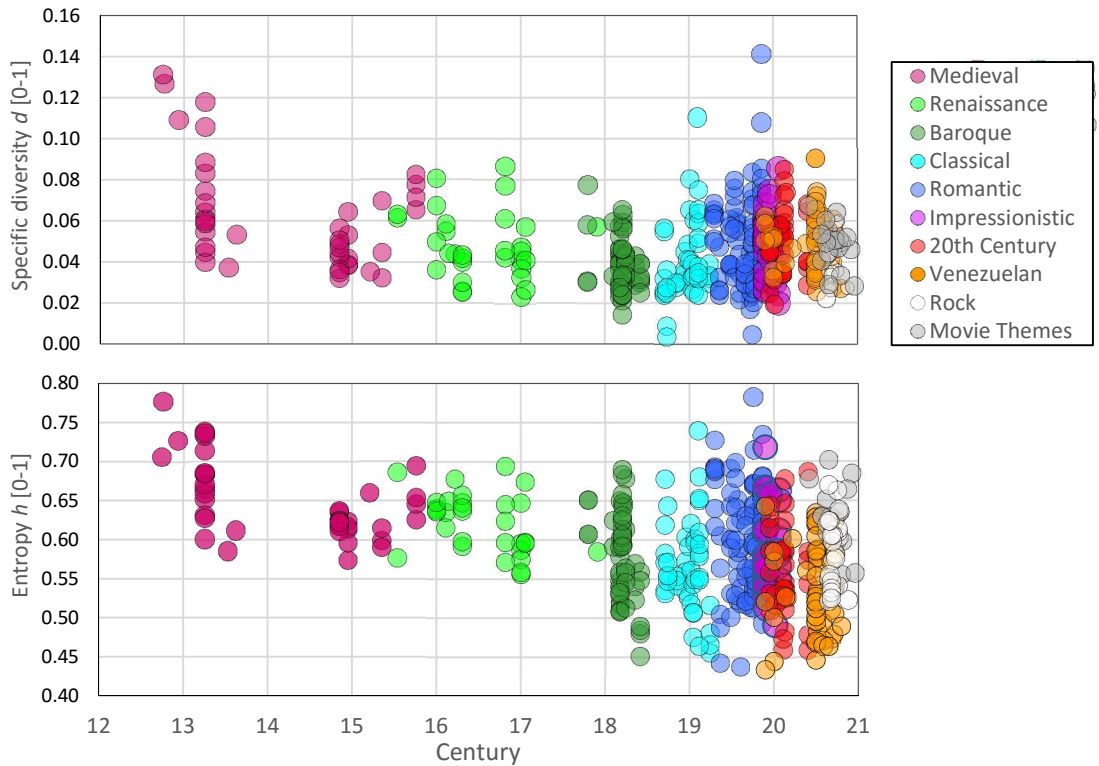

**Appendix J. Music styles by composer, in the space *(specific diversity, entropy, 2nd order entropy)*, ($d, h^{[1]}, h^{[2]}$).**

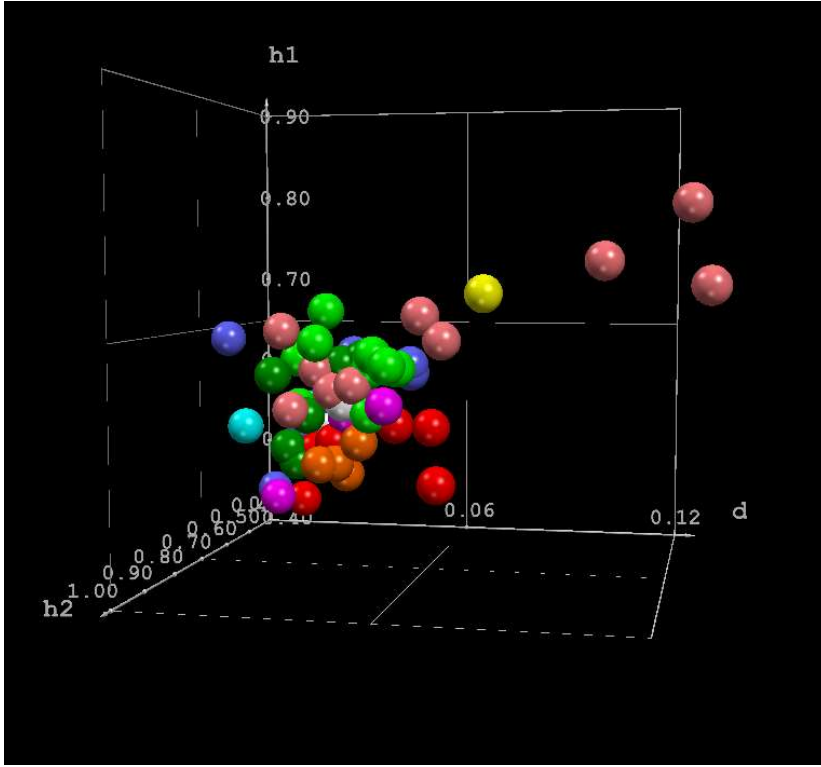

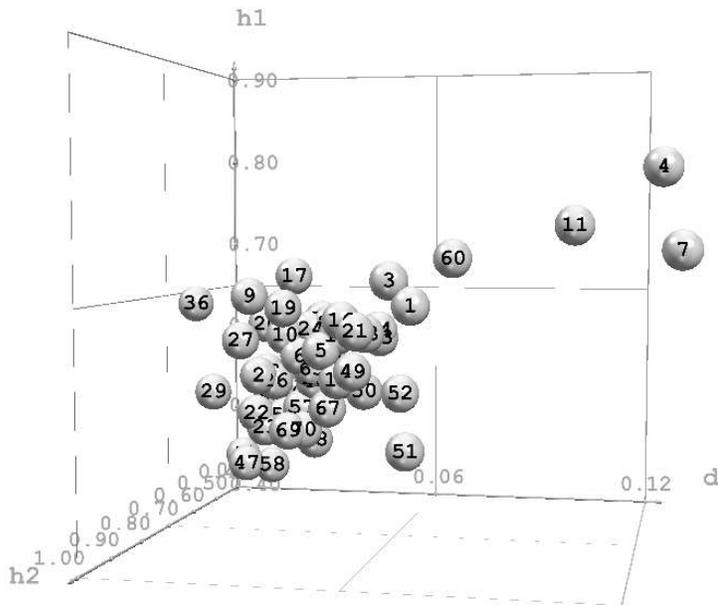

| | |
|---|---|
| 1 | AGRICOLA.Alexander |
| 2 | AlfonsoX.ElSabio |
| 3 | Anonimous |
| 4 | DeBORNEIL.Guiraut |
| 5 | DeLaHALLE.Adam |
| 6 | DePERUSTO.Matheus |
| 7 | DIE.Beatrice |
| 8 | DuFAY.Guillaume |
| 9 | DUNSTABLE |
| 10 | SOLAGE |
| 11 | VAQUEIRAS |
| 12 | Anonimous |
| 13 | BYRD.William |
| 14 | CAPRIOLA.Vincenzo |
| 15 | CLARK.Jeremiah |
| 16 | ENCINA.JuanDel |
| 17 | LEMLIN.Lorenz |
| 18 | MONTEVERDI.Claudio |
| 19 | SENFL.Ludwig |
| 20 | SUSATOTielman |
| 21 | OCKEGEHEM.Johannes |
| 22 | BACH.JohannSebastian |
| 23 | HANDEL.GeorgeFriedrich |
| 24 | PECHELBEL.Johann |
| 25 | SCARLATTI.Domenico |
| 26 | TELEMANN.Georg |
| 27 | VIVALDI.Antonio |
| 28 | HAYDN.FranzJoseph |
| 29 | Mozart |
| 30 | Beethoven |
| 31 | PAGANINI.Niccolo |
| 32 | SCHUBERT.Franz |
| 33 | BERLIOZ.Hector |
| 34 | BIZET.Georges |
| 35 | BORODIN.Alexander |
| 36 | Chopin |
| 37 | DVORAK.Antonin |
| 38 | GRIEG.Edvard |
| 39 | LIZT.Franz |
| 40 | MAHLER.Gustav |
| 41 | MUSSORGSKY.Modest |
| 42 | RIMSKY.KORSAKOV.Nikolai |
| 43 | SAINTSAENS.Camille |
| 44 | SMETANA.Bedrich |
| 45 | TCHAIKOVSKI.PeterIlich |
| 46 | DEBUSSY.Claude |
| 47 | GRIFFES.Charles |
| 48 | RAVEL.Maurice |
| 49 | SATIE.Erik |
| 50 | BARTOK.Bela |
| 51 | CASELLA.Alfred |
| 52 | CHANG.ChenKuang |
| 53 | COPLAND.Aaron |
| 54 | DeFALLA.Manuel |
| 55 | ELGAR.SirEdward |
| 56 | RACHMANINOV.Sergei |
| 57 | SHOSTAKOVICH.Dmitri |
| 58 | SIBELIUS.Jean |
| 59 | Chinese |
| 60 | Hindu-Raga |
| 61 | MovieThemes |
| 62 | Enya |
| 63 | EricClapton |
| 64 | Queen |
| 65 | RollingStones |
| 66 | TheBeatles |
| 67 | LAURO.Antonio |
| 68 | ROMERO.Aldemaro |
| 69 | DIAZ.Simon |
| 70 | VARIOUS.Composers |

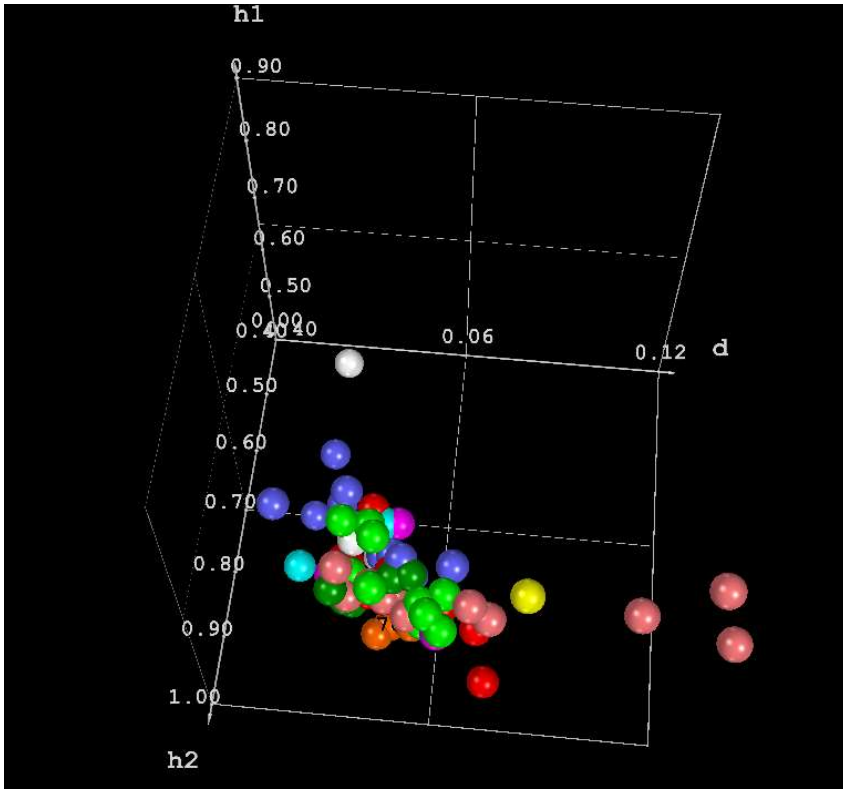
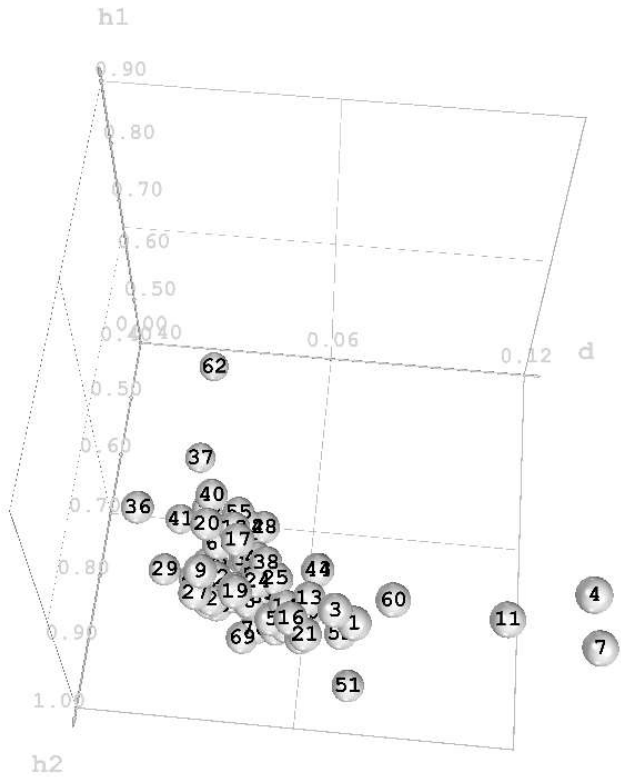

| 1 | AGRICOLA.Alexander |
|---|---|
| 2 | AlfonsoX.ElSabio |
| 3 | Anonimous |
| 4 | DeBORNEIL.Guiraut |
| 5 | DeLaHALLE.Adam |
| 6 | DePERUSTO.Matheus |
| 7 | DIE.Beatrice |
| 8 | DuFAY.Guillaume |
| 9 | DUNSTABLE |
| 10 | SOLAGE |
| 11 | VAQUEIRAS |
| 12 | Anonimous |
| 13 | BYRD.William |
| 14 | CAPRIOLA.Vincenzo |
| 15 | CLARK.Jeremiah |
| 16 | ENCINA.JuanDel |
| 17 | LEMLIN.Lorenz |
| 18 | MONTEVERDI.Claudio |
| 19 | SENFL.Ludwig |
| 20 | SUSATOTielman |
| 21 | OCKEGEHEM.Johannes |
| 22 | BACH.JohannSebastian |
| 23 | HANDEL.GeorgeFriedrich |
| 24 | PECHELBEL.Johann |
| 25 | SCARLATTI.Domenico |
| 26 | TELEMANN.Georg |
| 27 | VIVALDI.Antonio |
| 28 | HAYDN.FranzJoseph |
| 29 | Mozart |
| 30 | Beethoven |
| 31 | PAGANINI.Niccolo |
| 32 | SCHUBERT.Franz |
| 33 | BERLIOZ.Hector |
| 34 | BIZET.Georges |
| 35 | BORODIN.Alexander |
| 36 | Chopin |
| 37 | DVORAK.Antonin |
| 38 | GRIEG.Edvard |
| 39 | LIZT.Franz |
| 40 | MAHLER.Gustav |
| 41 | MUSSORGSKY.Modest |
| 42 | RIMSKY.KORSAKOV.Nikolai |
| 43 | SAINTSAENS.Camille |
| 44 | SMETANA.Bedrich |
| 45 | TCHAIKOVSKI.PeterIlich |
| 46 | DEBUSSY.Claude |
| 47 | GRIFFES.Charles |
| 48 | RAVEL.Maurice |
| 49 | SATIE.Erik |
| 50 | BARTOK.Bela |
| 51 | CASELLA.Alfred |
| 52 | CHANG.ChenKuang |
| 53 | COPLAND.Aaron |
| 54 | DeFALLA.Manuel |
| 55 | ELGAR.SirEdward |
| 56 | RACHMANINOV.Sergei |
| 57 | SHOSTAKOVICH.Dmitri |
| 58 | SIBELIUS.Jean |
| 59 | Chinese |
| 60 | Hindu-Raga |
| 61 | MovieThemes |
| 62 | Enya |
| 63 | EricClapton |
| 64 | Queen |
| 65 | RollingStones |
| 66 | TheBeatles |
| 67 | LAURO.Antonio |
| 68 | ROMERO.Aldemaro |
| 69 | DIAZ.Simon |
| 70 | VARIOUS.Composers |

**Appendix K. Average positon of the classical composers represented in the space specific diversity, entropy, 2nd order entropy ($d, h^{[1]}, h^{[2]}$).**

Each bubble represents the dominant location of the music of a composer.

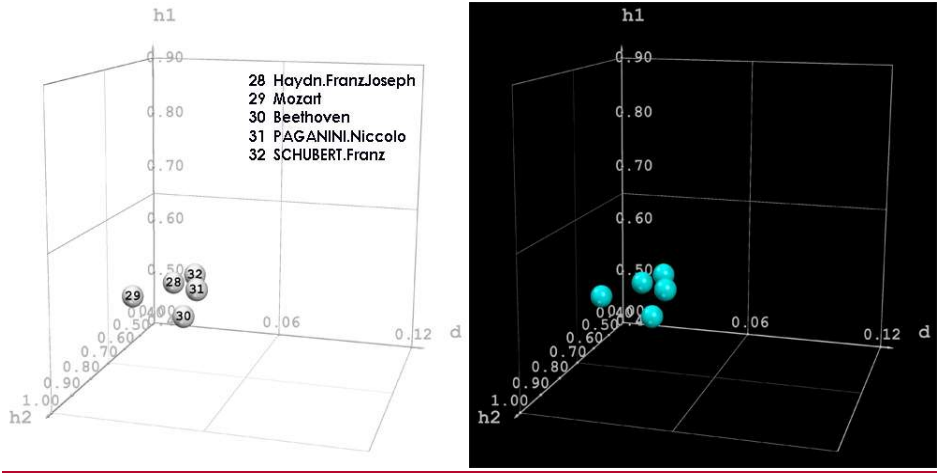